\documentclass[a4paper,usenatbib]{mn2e}
\usepackage[T1]{fontenc}
\usepackage{aecompl}
\usepackage{float}
\restylefloat{table}
\usepackage{times}
\usepackage{natbib}
\makeatletter
\renewcommand\hyper@natlinkbreak[2]{#1}
\makeatother
\usepackage{etoolbox}
\usepackage{caption}
\usepackage{subfigure}
 \usepackage{graphicx}
 \usepackage{graphics}
 \usepackage{amsmath}
  \usepackage{lipsum} 
 \usepackage{afterpage}
 \newcommand{\hi}{\mbox{H{\sc i}}}

\newcommand{\ha}{H$\alpha$}

\newcommand{\kms}{km s$^{-1}$}

 \renewcommand\appendix{\par
  \setcounter{section}{0}
  \setcounter{subsection}{0}
  \setcounter{figure}{0}
  \setcounter{table}{0}
  \renewcommand\thesection{ \Alph{section}}
  \renewcommand\thefigure{\Alph{section}\arabic{figure}}
  \renewcommand\thetable{\Alph{section}\arabic{table}}
}

\title[KAT-7 observations of NGC 253]{$\hi$ observations of the nearest starburst galaxy NGC 253\\
with the SKA precursor KAT-7}
\author[Lucero et al.]{D. M. Lucero$^{1}$\thanks{dlucero@ast.uct.ac.za}, C. Carignan$^{1}$, E. C. Elson$^{1}$, T. H. Randriamampandry$^{1}$, T. H. Jarrett$^{1}$,
\and T. A. Oosterloo$^{2,3}$ and G. H. Heald$^{2,3}$ \\
\\
$^{1}$ Department of Astronomy, University of Cape Town, Private Bag X3, Rondebosch 7701, South Africa\\ 
$^{2}$ Netherlands Institute for Radio Astronomy (ASTRON), Postbus 2, 7990 AA Dwingeloo, The Netherlands\\
$^{3}$ Kapteyn Astronomical Institute, University of Groningen, PO Box 800, 9700 AV Groningen, The Netherlands\\}

\pagerange{\pageref{firstpage}--\pageref{lastpage}} \pubyear{2015}

\begin{document}

\date{}

\label{firstpage}
\maketitle

\begin{abstract}
We present \hi\ observations of the Sculptor Group starburst spiral galaxy NGC 253, obtained with the Karoo Array Telescope (KAT-7). KAT-7 is a pathfinder for the SKA precursor MeerKAT, under construction. The short baselines and low system temperature of the telescope make it very sensitive to large scale, low surface brightness emission. The KAT-7 observations detected 33\% more flux than previous VLA observations, mainly in the outer parts and in the halo for a total \hi\ mass of $2.1 \pm 0.1$ $\times 10^{9}$ M$_{\odot}$.  \hi\ can be found at large distances perpendicular to the plane out to projected distances of $\sim$9-10 kpc away from the nucleus and $\sim$13-14 kpc at the edge of the disk.  A novel technique, based on interactive profile fitting, was used to separate the main disk gas from the anomalous (halo) gas. The rotation curve (RC) derived for the HI disk confirms that it is declining in the outer parts, as seen in previous optical Fabry-Perot measurements. As for the anomalous component, its RC has a very shallow gradient in the inner parts and turns over at the same radius as the disk, kinematically lagging by $\sim$100 km/sec. The kinematics of the observed extra planar gas is compatible with an outflow due to the central starburst and galactic fountains in the outer parts. However, the gas kinematics shows no evidence for inflow. Analysis of the near-IR WISE data, shows clearly that the star formation rate (SFR) is compatible with the starburst nature of NGC 253.
\end{abstract}


\section{Introduction}

The seven-dish KAT-7 array \citep{car13} was built as an engineering testbed for the 64-dish Karoo Array Telescope, known as MeerKAT, which is the South African precursor of the Square Kilometre Array (SKA).  KAT-7 and MeerKAT are located close to the South African SKA core site in the Northern Cape's Karoo desert region. Construction of the array was completed in December 2010. The array is extremely compact, with baselines ranging from 26 m to 185 m and the receivers have a very low $T_{sys} \sim 30$K.  While its main purpose is to test technical solutions for MeerKAT and the SKA, scientific targets, such as NGC 253, were also observed during commissioning to test the \hi\ line mode.  In this paper, we present over 150 hours of observations taken with KAT-7 in order to study its large scale extra-planar \hi\ gas component.

NGC 253 is the nearest example of a galaxy undergoing enhanced star formation in the South, M82 being the nearest one in the North. Its nucleus contains a starburst with a star-formation rate (SFR) of $\sim$5 M$_{\odot}$ yr$^{-1}$, roughly 70\% of the rate of the entire galaxy \citep{wik14}. It is considered a prototype starburst galaxy \citep[see e.g.][]{sak11, str02} and \citet{rie88} suggest that it may be in an earlier starburst phase than M82.  Of particular interest is that, despite the high SFR, NGC 253 is a non-interacting (non-merger) system, which does not exclude that it occurred in the past (see Sec. \ref{sec:starburst}).

\begin{figure}
\centering
\begin{tabular}{cc}
\includegraphics[width=\columnwidth]{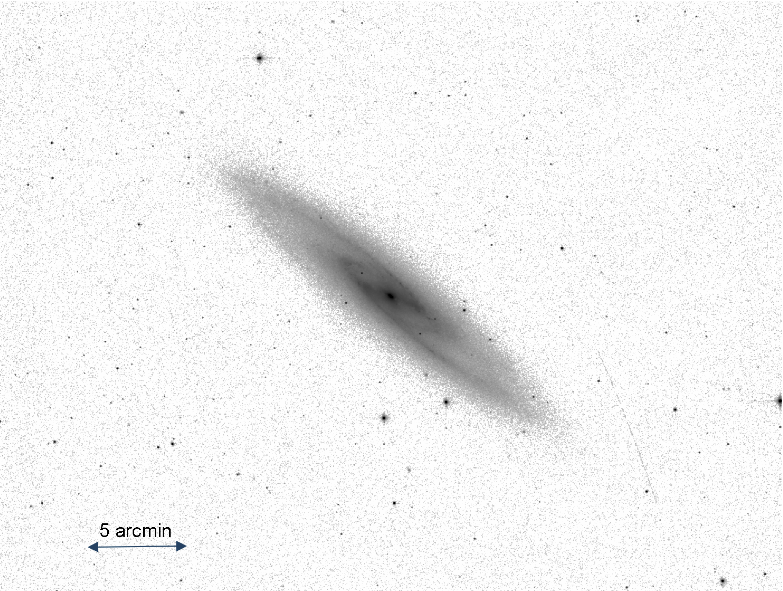}\\
\includegraphics[width=\columnwidth]{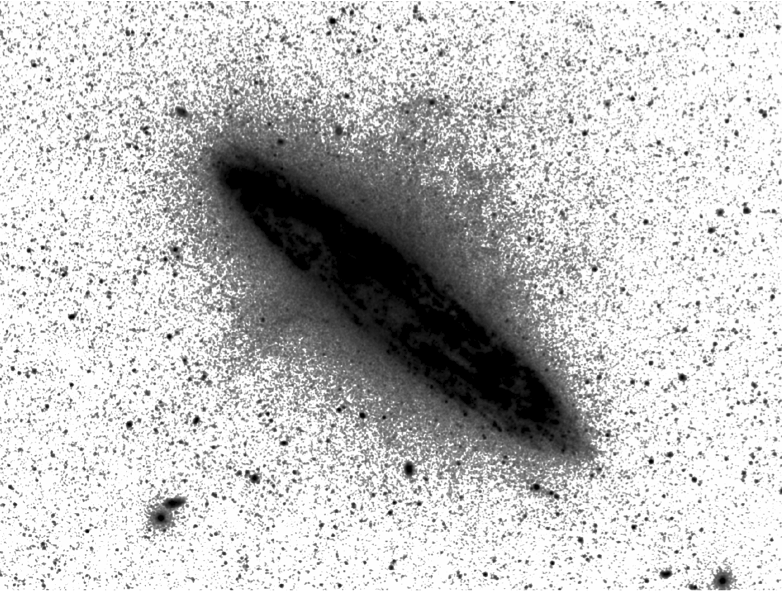}\\
\end{tabular}
\caption{2MASS K-band (top) and GALEX FUV (bottom) images of NGC 253.}
\label{fig:FUV}
\end{figure}

One very interesting feature is the diffuse extra-planar \hi\ surrounding the X-ray emission, first seen in the ATCA observations of \citet{boo05}.  Not only is the nuclear starburst forming stars at a fairly high rate but it is also thought to produce a super-wind \citep{hec90}, which expels material into the halo.  The high sensitivity of KAT--7 to large scale diffuse emission  \citep{car13} allows us to better study the known extra-planar gas component. Such extra-planar gas is expected when looking at a deep UV image showing emission far from the plane, compared to a 2MASS image \citep{jar03}, which shows the main stellar disk component (Figure \ref{fig:FUV}).

In NGC 253, most of the activity is confined within $\sim$1 kpc of the nucleus, fuelled by bar-driven dense cold molecular gas that obscures the starburst \citep{rie80, hou97, eng98, sak11}. Recent X-ray, optical and ALMA mm-wave observations trace outflowing gas that is powered by the starburst \citep{mit13, bol13}.  Owing to proximity, the two nearest starbursts, NGC 253 and M82, are ideal laboratories for detailed (pc-scale) study of nuclear outflows and galactic-disk fountains. Conversely, their global properties -- which place the starbursts in context with the total gas and stellar content -- are a challenge to measure due to their large angular size and brightness (frequently saturating detectors).  While large angular scales are not a problem at optical wavelengths, for radio interferometry imaging studies, one needs the proper short baselines to be able to detect large structures and the compactness of KAT-7 makes it better suited than e.g. the VLA for the observation of those large scales in nearby objects (see Table \ref{LAS}).


Recent deep \hi\ observations of several nearby galaxies indicate that up to 15\% of the neutral hydrogen of a spiral galaxy is located in the halo \citep{hea14}. The gas outside the main \hi\ disk of spirals has been studied in many galaxies: NGC 891 \citep{swa97, oos07}, NGC 2403 \citep{sch00, fra01}, NGC 2997 \citep{hes09}, NGC 3198 \citep{gen13}, 
NGC 4244 \citep{zsc11}, NGC 4559 \citep{bar05}, NGC 4565 \citep{zsc12}, NGC 5775 \citep{lee01}, NGC 6946 \citep{boo08},
M31 \& M33 \citep{wei05, thi04} and the Milky Way (MW) through High Velocity Clouds (HVCs) \citep{wak97}, the Magellanic Stream \citep{mat74} and the Leading Arm \citep{put98}. See also  \citet{hea11, hea14} for a description of the HALOGAS survey and \citet{san08, put12} for reviews.

The interaction between the gas in the halo and the disk of spirals is believed to play an important role in their evolution. Halo gas connects the baryon-rich intergalactic medium (IGM) to the star-forming disks of galaxies \citep{put12}. The classical "galactic fountain" scenario \citep{sha76}, where hot gas is expelled from the disk through "galactic chimneys" created by multiple supernova explosions from clusters of young massive stars \citep{nor96} 
can explain the presence of hot gas in the halo. It is believed that, as the gas expands, it eventually converts back to \hi\ as it cools through radiative losses; raining back down on the disk \citep{bre80} and feeding subsequent star formation \citep[see also the models for the MW:][]{mar12}.  A good example of such exchanges between the disk and the halo is the \ha\ kinematical study by  \citet{cec01} of one of the filament in the super wind outflow of NGC3079.  Long ago,  \citet{oor66} also suggested an external origin for some of that gas in the form of primordial gas clouds left over from the formation of the galaxies. Finally, as suggested by the Magellanic Stream and the Leading Arm in the MW, some of that halo gas could come from ISM torn out of dwarf galaxies during close encounters with a large spiral.

\begin{figure}
\includegraphics[width=\columnwidth]{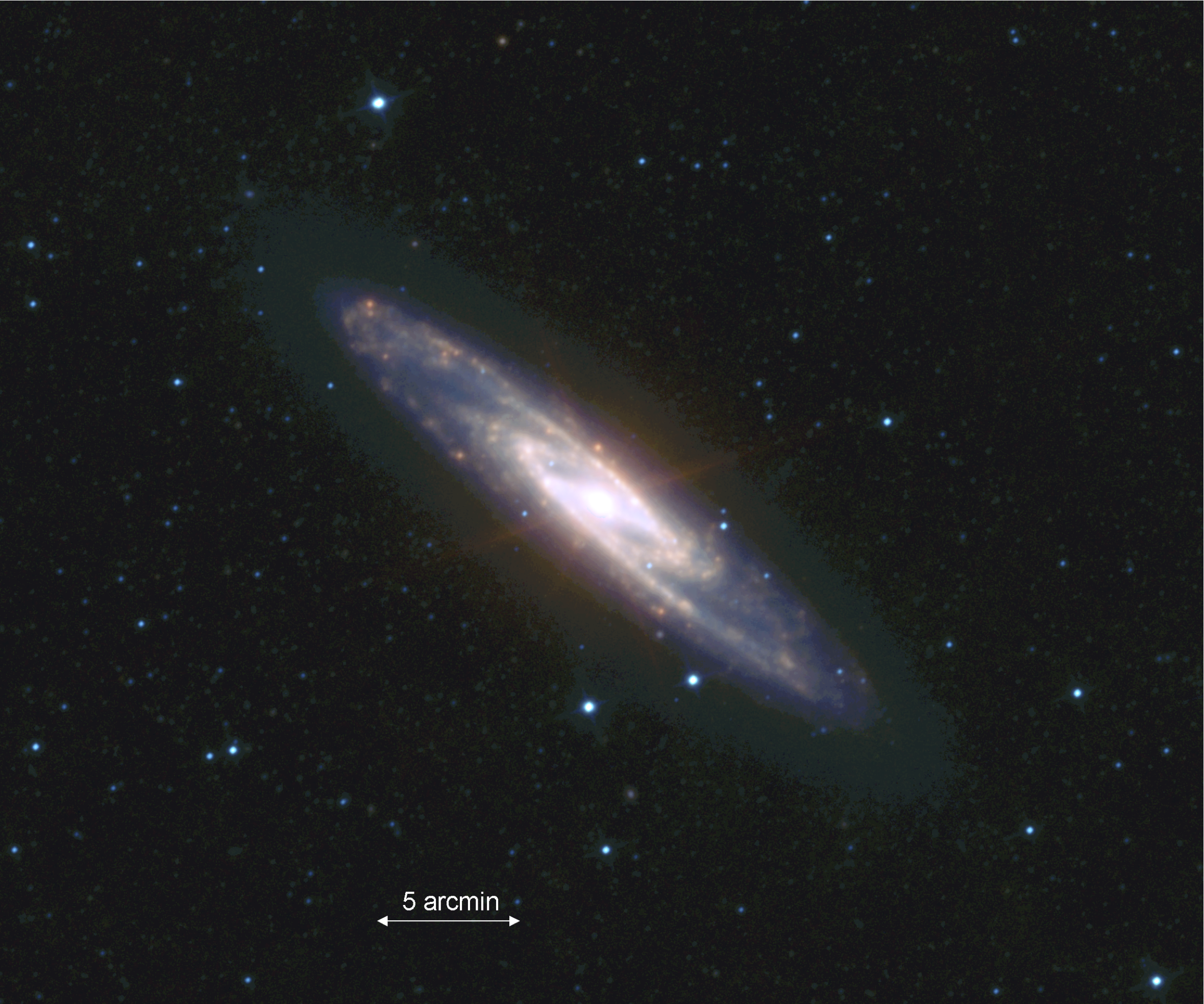}
\caption{Wise W1 (blue), W2 (green), W3 (orange) and W4 (red) color-combination image of NGC 253.}
\label{fig:WiseZoom}
\end{figure}

The optical and IR parameters of NGC 253 are summarized in Table \ref{optpar}. It is the brightest and earliest type among the five late-type spirals composing the nearest group to us, the Sculptor group \citep{puc88}. 
It is classified SAB(s)c  by \cite{dev91} but, as can be seen in Figure \ref{fig:WiseZoom}, it is clearly a barred SB(r)c galaxy on IR images. The different recent distance estimates from the Planetary Nebulae Luminosity Function (PNLF) and the Tip of the Red Giant Branch (TRGB) distance indicators \citep{rek05, mou05, del09, jac09, his11}
give a distance of 3.5 Mpc, for a scale of $\simeq$1 kpc per arcmin. While NGC 55 and NGC 300 are on the near side and NGC 7793 on the far side of the Sculptor group, NGC 253 is at the mean distance of the group with NGC 247, which is only at a projected distance of $\sim$350 kpc to the North. 

\begin{table}
\centering
\caption{Optical \& IR (WISE) parameters of NGC 253.}
\label{optpar}
\begin{tabular}{lr}
\hline\hline
Parameter  & Value\\          
\hline
Morphological type$^1$                                  	& SB(r)c       \\
Right Ascension$^2$  (J2000)                          	&0$^{\rm h}$ 47$^{\rm m}$ 33.1$^{\rm s}$      \\
Declination$^2$  (J2000)                                  	& -25$^{\rm o}$  17\arcmin\ 18\arcsec     \\
Distance (Mpc)                                                 	& 3.47 $\pm\ 0.14$                            \\
Scale (kpc arcmin$^{-1}$)                              	& $\sim 1$    \\
Isophotal major diameter$^2$, $D_{25}$  		& 27.5\arcmin \\
Holmberg radius$^3$, $R_{HO}$ 			& 17.5\arcmin \\
Axis ratio$^2$ (q=b/a), $R_{25}$ 			& 0.25 \\
Inclination (q$_0$=0.15) 					& 78.3$^0$ \\
Position Angle							& 52$^0$ \\
Total  apparent $B$ magnitude$^2$			& 8.04  \\
Corrected apparent $B$ magnitude$^2$         	&  7.02  \\
Absolute $B$ magnitude                                    	&  -20.68           \\ 
Total B luminosity, $L_{\odot}$ 				& $2.78 \times 10^{10}$ \\
Absolute $W1$ magnitude, WISE			&-24.24  \\
Color (W1 - W2), WISE					&0.21 \\
Log(M$_*$), WISE, $M_{\odot}$			&10.33  \\
\hline
(1) See Figure \ref{fig:WiseZoom}\\
(2) \citet{dev91}\\
(3) Radius at $\mu_B$ = 26.6 mag arcsec$^{-2}$\\
\end{tabular}
\end{table}

The first detailed $\hi$ observations of this galaxy \citep{huc72, com77} show clearly an asymmetric $\hi$ distribution.
The presence of a strong continuum source in the central regions \citep{hum84a} suggests that any kinematical analysis in the very inner parts
is to be undertaken with great care. From the low spatial resolution single dish early $\hi$ observations, the rotation curve was believed to be declining at large radii
\citep{huc75} but higher resolution VLA observations \citep{puc91} show a rotation curve still rising at the last observed velocity point ($\sim$12\arcmin). 
However, recent Fabry-Perot observations \citep{bfq97, hla11} detected \ha\ emission further out than previous \hi\ observations. While
the \ha\ kinematics agree with the \hi\ for R $\leq$ 12\arcmin, it suggests a declining rotation curve between 15\arcmin and 19\arcmin. 
It will be interesting to see if our KAT-7 observations extend to those radii and confirm this decline.

\begin{figure}
\centering
\begin{tabular}{ccc}
\includegraphics[width=80mm]{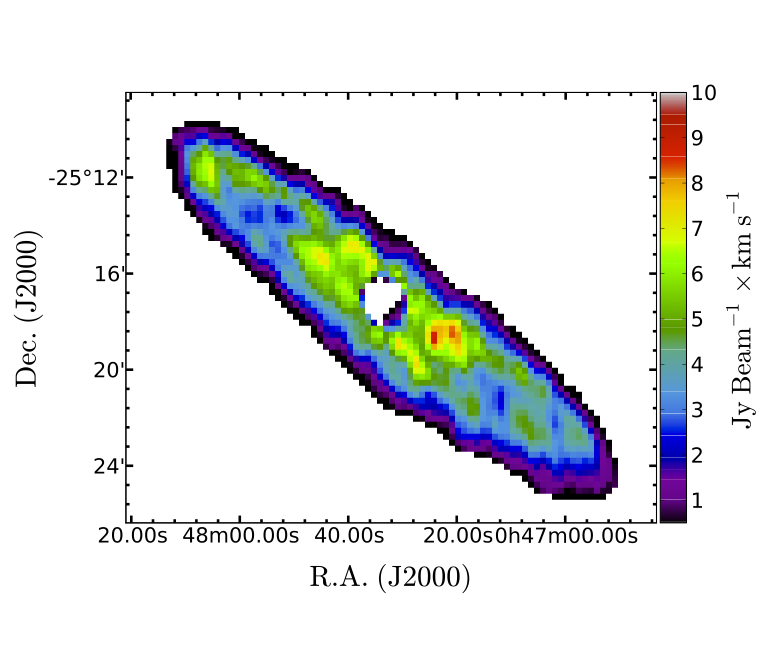}\\
\includegraphics[width=80mm]{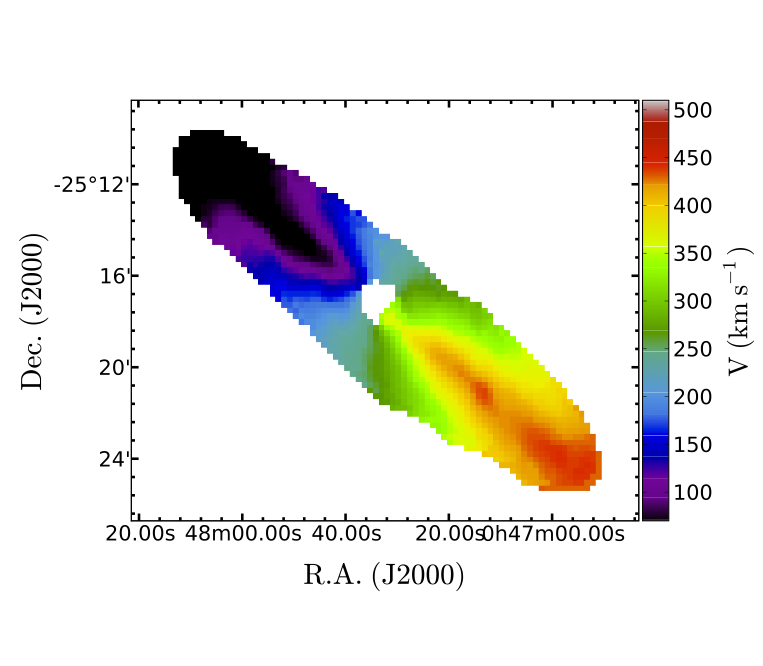}\\
\includegraphics[width=80mm]{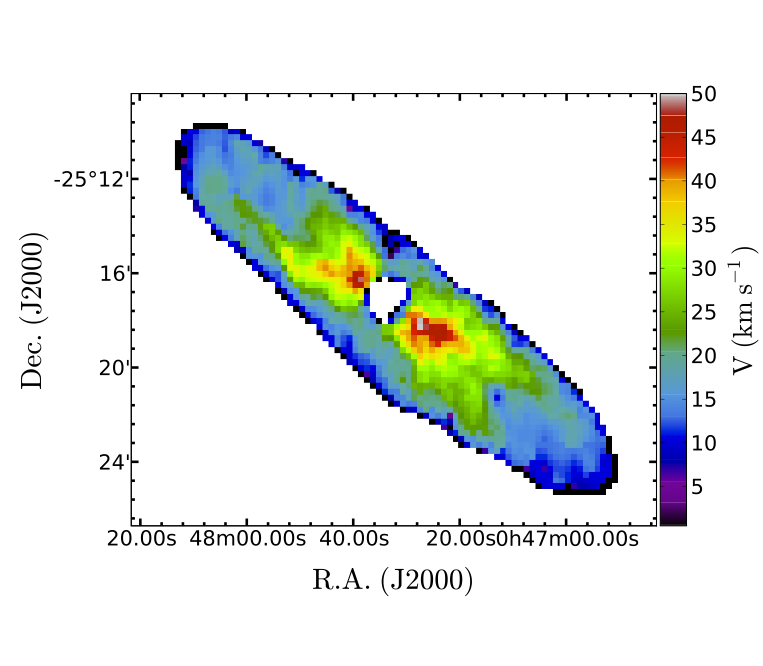}\\
\end{tabular}
\caption{VLA moment maps (top: \hi\ distribution; middle: velocity field; bottom: velocity dispersion) of NGC 253 derived from the data set presented by \protect\citep{puc91}. The velocity resolution is 20.6 \kms\ and the spatial resolution is $68\arcsec \times 68\arcsec$.}
\label{fig:VLA}
\end{figure}

Figure \ref{fig:VLA} shows previous higher spatial resolution ($68\arcsec \times 68\arcsec$ vs $213\arcsec \times 188\arcsec$) VLA observations \citep{puc91}. Note however that the final KAT-7 data have a factor of $\sim$4 better velocity resolution. Regarding the \hi\ distribution, the first feature to notice is the strong absorption in the centre, 
which prevents any reliable kinematics to be extracted close to the starburst region. It can be seen that besides tracing the spiral arms, the \hi\ is really concentrated around the nucleus. As for the velocity field, it is a textbook case in terms of regularity. Only a slight increase in position angle is suggested in the very outer parts. The most interesting map is the velocity dispersion map, which shows a clear increase  $\sim$45-50 \kms\ around the starburst region while it is fairly constant $\sim$10-15 \kms\ in the rest of the disk. Because of the short exposure time ($\sim$30 minutes in D configuration) and the lack of short spacings, those observations only show the disk emission and were insensitive to the large scale ($>$ 15\arcmin) low column density extra planar \hi.

In theory, the Largest Angular Sizes (LAS) accessible to an aperture synthesis telescope is given by:
\begin{equation}
 \rm {LAS = 0.6 \times \lambda / B_{min}}
\end{equation}

When trying to observe LAS, Table \ref{LAS} illustrates why KAT-7 should be a good instrument for this purpose. If we consider the shortest physical baseline, it can be seen that the gain of KAT-7 is $\sim$33\% over the VLA, WSRT (in its Maxishort configuration) and ATCA. In fact, the minimum projected baseline is the antenna diameter. This means that if we instead consider the shadowing limit, the gain over the VLA, WSRT and ATCA is a factor of 2 ($\sim$ 36\arcmin\ vs 18\arcmin).

\begin{table}
\centering
\caption{Largest Angular Sizes (LAS) of different synthesis telescopes. Given for the different telescopes (1) are: minimum baseline (2), 
antenna diameter (3), LAS from the minimum physical baseline (4), and LAS down to the shadowing limit (5). }
\label{LAS}
\begin{tabular}{ccccc}
\hline\hline
 (1) & (2) & (3) & (4) & (5)  \\
Telescope & Min. bl 	& Ant. diam.  & LAS & LAS \\
	        &  (m)  	& (m) & (min. bl) & (sha. lim.)  \\
\hline
KAT-7	 	& 26 		& 12	& 16.7\arcmin	& 36.1\arcmin \\
VLA-D		& 35		& 25	& 12.4\arcmin 	& 17.3\arcmin  \\
WSRT-MS	 	& 36		& 25	& 12.0\arcmin 	& 17.3\arcmin  \\
GMRT		& 100	& 45 	& 4.3\arcmin 	& 9.6 \arcmin \\
ATCA		& 31		& 22 	& 14.0\arcmin 	& 19.7\arcmin \\
\hline
\end{tabular}
\end{table}

In practice, the largest scale that can be reliably imaged cannot be assessed as simply as Table~\ref{LAS} suggests. When less flux is recovered with an interferometer than with a single dish, the shortcoming is often explained by a lack of short spacings. Yet it may often be the case that additional care in calibration and imaging can resolve the discrepancy. The resulting quality when imaging large-scale emission is critically dependent on several factors including the image reconstruction techniques that are employed (classic {\tt CLEAN}, MEM, multi-resolution {\tt CLEAN}, etc). It is important to note that the spatial scales of interest should be assessed per spectral channel, not for the entire galaxy. In single channels, the angular scales are always more restricted. A good illustration of imaging structures larger than the nominal LAS is the study of NGC 628 by \citet{kam92}, in which features as large as $\sim$25\arcmin are seen in individual channels (see their Fig. 1 showing the naturally weighted channel maps), although emission on scales larger than $\sim$17.5\arcmin should not be seen by the VLA in D configuration. Nevertheless, it remains true that KAT-7's smaller dish diameter and shorter baselines enable it to image larger structures than with the VLA, WSRT and ATCA.

The remainder of this paper is as follow.  In Sec.~\ref{sec:ODR}, a description of the KAT-7 observations and data reduction will be given. Sec.~\ref{sec:continuum} will discuss the continuum emission and Sec.~\ref{sec:HICD} will describe the \hi\ content and distribution.  Sec.~\ref{sec:expl} will explain how to distinguish between the \hi\ in the disk and the \hi\ in the halo while Sec.~\ref{sec:VFRC} will study the \hi\ kinematics of both components. The main results will be discussed in Sec.~\ref{sec:dis} and,
finally, a summary and final conclusions will be given in Sec.~\ref{sec:con}.

\section{KAT-7 Observations and Data Reduction}
\label{sec:ODR}

The KAT-7 dishes have a prime-focus alt-az design with a F/D of 0.38,
optimized for single-pixel L-band feeds. The low noise amplifiers
(LNAs) for  the feeds are cryogenically cooled to 80 K using Stirling
coolers. The key system specifications for KAT-7 are summarized in
Table~\ref{K7spec}. The digital backend of the system uses the
Reconfigurable Open Architecture Computing Hardware 
 (ROACH\thanks{https://casper.berkeley.edu/wiki/ROACH}), which is a
flexible and scalable system enabling spectral line modes covering a
wide range of resolutions. Table~\ref{K7corr} gives the details of
the recently commissioned correlator modes. Digital filters give a
flat bandpass over the inner 75\% of the band with a rapid roll-off at
the edges of the band.

\begin{table}
\centering
\caption{KAT--7 specifications.}
\label{K7spec}
\begin{tabular}{lr}
\hline\hline
Parameter & Value \\
\hline
Number of antennas 		& 7\\
Dish diameter			& 12 m \\
Min baseline			& 26 m\\
Max baseline			& 185 m\\
Frequency range		& 1200 - 1950 MHz \\
Max instantaneous bandwidth & 256 MHz \\
Polarisation 			& Linear H \& V \\
T$_{\rm sys}$			& 26 K \\
Aperture efficiency   		& 0.65 \\
System Equivalent Flux Density & 1000 Jy \\
Latitude				&  -30:43:17.34 \\
Longitude				& 21:24:38.46 \\
Elevation				& 1038 m \\
Digital back-end		& ROACH boards\\
\hline
\end{tabular}
\end{table}

\begin{table}
\centering
\caption{KAT-7 correlator modes.}
\label{K7corr}
\begin{tabular}{lccc}
\hline\hline
mode & total BW & number of  & channel BW \\
		&	(MHz)				& 		channels				& (kHz) \\
\hline
c16n2M4k & 1.5625		& 4096							& 0.381 \\
c16n7M4k & 6.25			& 4096							& 1.526 \\
c16n25M4k & 25			& 4096							& 6.104 \\
c16n400M4k & 256		& 1024 					& 390.625 \\
\hline
\end{tabular}
\end{table}

The parameters of the KAT-7 observations of NGC 253 are given in Table~\ref{kat7par}.
The data were collected over 17 observing sessions between 2013 March and 2013 July using the c16n25M4K spectral line mode.  This correlator mode gives velocity channels of 1.3 km s$^{-1}$ over a flat bandpass of ~4000 km s$^{-1}$.  The larger bandwidth was used to look for background galaxies in the field and to ensure a good estimation of the continuum for subtraction.  All antennas were in the array during all the observing sessions.  The median time of each session on target was 9 hours for a total project time of 153.5 hours, including calibration and slew time.  Total time on NGC 253 was 115.6 hours.  

\begin{table}
\centering
\caption{Parameters of the KAT--7  observations.}
\label{kat7par}
\begin{tabular}{lr}
\hline\hline
Parameter & Value \\
\hline
Start of observations & 5 March 2013 \\
End of observations & 19 July 2013 \\
Total integration (on source) &  115.6 hours\\
FWHM of primary beam & 1.08\degr \\
Total Bandwidth & 25 MHz \\
Central frequency &  1419 MHz \\
Channel Bandwidth & 6.1 kHz \\
Number of channels & 4096 \\
Channel width & 1.28 \kms \\
Maps gridding & 20\arcsec\ x 20\arcsec \\
Maps size & 257 x 257 \\
Flux calibrator	& 3C138 \\
Phase/bandpass calibrator  & 0023-263 \\
\hline
Robust = 0 weighting function & \\
FWHM of synthesized beam & 213\arcsec  x 188\arcsec \\
RMS noise  & 1.0 mJy/beam \\
Column density limit (3$\sigma$)  & $1.3 \times 10^{19}$ cm$^{-2}$ \\
\hline
\end{tabular}
\end{table}

Each of the seventeen observing sessions were reduced separately.  All data calibration was done using standard calibration tasks in the Common Astronomy Software Applications (CASA 3.4.0) 
package \citep{mcm07}.  The KAT-7 primary beam is large enough to observe NGC 253 in a single pointing.  Phase drifts as a function of time were corrected by means of a nearby point source (0023-263) observed every ten minutes.  This source was also used to correct for variations in the gain as a function of frequency (bandpass calibration).  The absolute flux scale was set by observations of 3C138.  Comparisons of the flux measurements on the observed calibrators suggests that the absolute flux uncertainties are on the order of 5\%.  Variations in the bandpass are on the order of 1\%.  

Initial imaging revealed which channels were free of HI emission.  Continuum emission was subtracted from the raw UV data by making first order fits to the line free channels using the CASA task UVCONTSUB.  The calibration was then applied and NGC 253 was then SPLIT from the calibration sources.  In order to make sure no residual calibration errors remained in the data, a test cube was made using the CASA task CLEAN.   KAT-7 does not employ doppler tracking and CASA does not fully recognize frequency keywords, so special care was taken to produce uv data sets and test cubes with the proper velocity coordinates \citep[see][]{car13}. 
The individual calibrated continuum subtracted uv data sets were then combined together using the CASA task CONCAT.  

Preliminary imaging of the combined data in CASA revealed the presence of artifacts in the form of horizontal lines which we were able to identify as the accumulation of RFI at u=0. This is most likely due to the fact that for those visibilities the fringe rotation is zero. These were removed by flagging all visibilities near u = 0. 
This uncovered further artifacts in the form of diagonal lines parallel to the major axis of NGC 253.  To remove this problem,  it was found necessary to  use the task SELFCAL in Miriad \citep{sau95}, on the strong continuum of NGC 253. This removed most of these artifacts giving a final rms of 1 mJy/beam with channels smoothed to 5 \kms\ in the final cube. 

\section{Continuum emission}
\label{sec:continuum}

\begin{figure}
\centering
\includegraphics[width=\columnwidth]{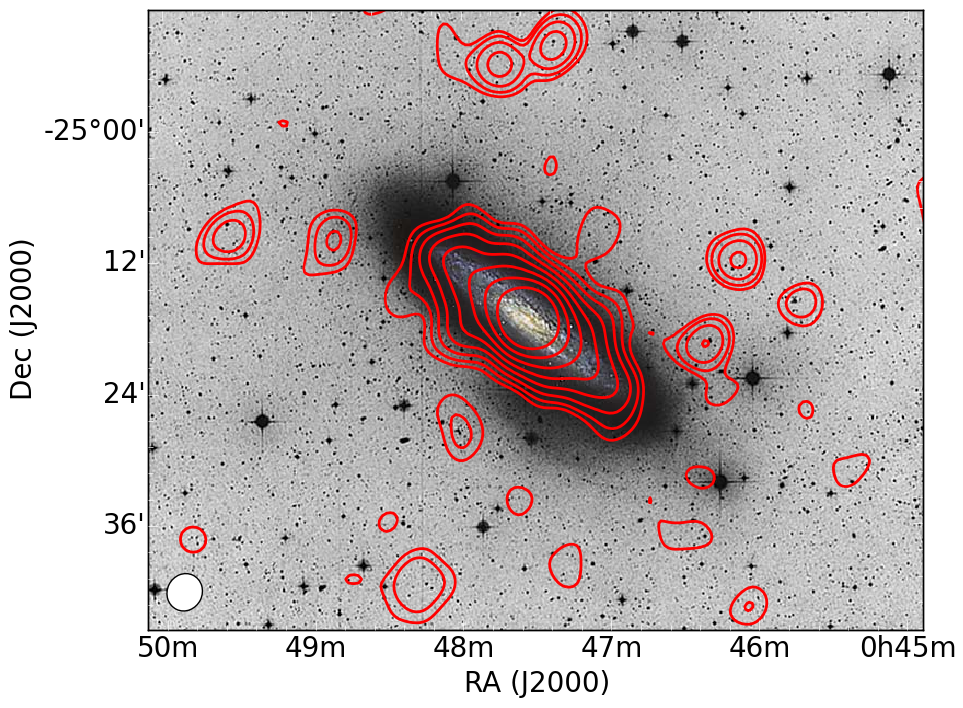}
\caption{Continuum map of NGC 253, overlaid on a deep optical image \protect\citep{mal97}. The contours are  7.5  ($3 \sigma$), 15, 30, 60, 120, 240, 480, 960  mJy/beam. The beam is in the bottom left corner.}
\label{fig:continuum}
\end{figure}

Figure \ref{fig:continuum} shows the KAT-7 continuum map of NGC 253. The total flux at 1.419 GHz is $6.8 \pm 0.3$ Jy. This is comparable with the previous measurements  of $6.5 \pm 0.6$ Jy at 1.415 GHz \citep{bea68} and $6.7 \pm 0.7$ Jy at 1.430 GHz \citep{hei63}. When comparing Fig. \ref{fig:continuum} to the observations at 1.46 GHz of \citet{hum84a} 
or at 1.49 GHz of \citet{con87}, 
the emission is slightly more extended ($\sim$20\%) along the major axis but 60\% more along the minor axis \citep[see also][]{car92}. In fact, the emission reaches the very edge of the optical disc on this very deep image from \citet{mal97}. However, the difference between our continuum map and the one of \citet{hum84a} is not surprising since they themselves mentioned that their emission is sitting in the middle of a negative bowl, a signature of the lack of short enough baselines at the VLA. 

Looking at the very high flux in the centre, it is clear that the starburst is responsible for a large part of the emission, while the spiral arms are more responsible for the lower flux in the outer disk. Along the major axis, it can be seen that the emission stops at the end of the spiral arms. Contrary to what is seen in the observations of some edge-on galaxies,  where the axis ratio of the continuum emission is similar to the axis ratio of the optical disk \citep{hum84b}, the axis ratio is twice that of the optical disk (0.50 vs 0.25) in the case of NGC 253. 

However, NGC 253 is not the only exception. NGC 4631 \citep{eke77}, which has strong X-ray emission originating from the nucleus and is interacting with a dwarf companion along the minor axis, also has a much thicker radio disk. This is only beaten by NGC 253's twin M82 \citep{sea91}, which has an axis ratio $\sim$1.0 (vs optical $\sim$0.4). Higher resolution (VLA, A configuration) 20cm observations of the centre of NGC 253 \citep{ulv97} show an even thicker disk with an axis ratio $\sim$0.65.

\citet{all78} explain that the sources of relativistic electrons are distributed like the light in the optical disk and that the electrons diffuse outwards. However, the larger thickness of the radio disk in this case suggests that the starburst may be responsible for an extra z-component of the magnetic field, which enhances the propagation of the relativistic electrons in the direction perpendicular to the plane. As discussed by \citet{hea12}, a substantial vertical displacement of an initially plane-parallel-ordered magnetic field may be driven by a star formation event. A detailed study of the magnetic field in the nuclear outflow of NGC 253 can be found in \citet{hee11}.

Since the KAT-7 observations were carried out over several months, we have a unique opportunity to investigate whether or not the continuum in the central regions exhibits time variability.  To this end we calculate the continuum flux inside a 2$\arcmin$ radius covering the absorption feature for all seventeen separate data sets.  No significant variation in continuum flux is observed.  All measured fluxes are within the measurement uncertainties ($\sim$ 5 \%).  It should be noted here again that the bulk of the radio continuum comes from the central starburst, reducing the likelihood of detecting significant variations in continuum flux.

\section{$\hi$ Content and Distribution}
\label{sec:HICD}

\begin{figure*}
\centering
\includegraphics[width=150mm]{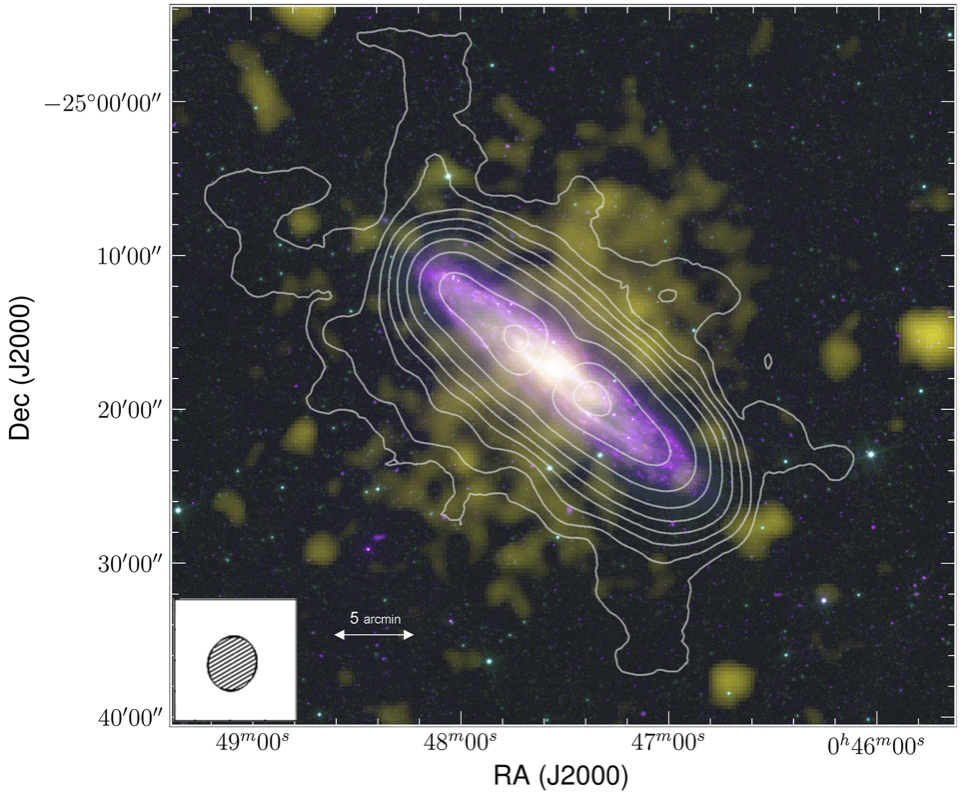}
\caption{Combined WISE (RGB) + FUV (magenta) + X-rays (yellow: Rosat) + HI contours of the moment 0 map (scale: 1\arcmin $\sim$ 1 kpc). The contours correspond to 1.3, 2.6, 5.2, 10.4, 20.8, 41.6 ,83.2, 166.4 \& 253.3  $\times 10^{19}$cm$^{-2}$. The ROSAT data are from \protect\citet{pie00}.}
\label{fig:WUVXdiskHI}
\end{figure*}

The total \hi\ distribution map, shown in Figure \ref{fig:WUVXdiskHI}, was derived using the task MOMNT (moment 0) in AIPS \citep{gre03}. It is superposed on a composite image of FUV from Galex, IR from WISE and soft X-rays from ROSAT (0.1-0.4 keV). The faintest level goes down to $\sim$1.0 $\times$ 10$^{19}$ cm$^{-2}$. At that level, the galaxy has a diameter of 34 $\pm 2$ kpc (the error being defined by the beam size), which is comparable to the optical diameter (cf. Tab. \ref{optpar}). It can be seen that a significant fraction of the  \hi\ can be found away from the plane out to projected distances of $\sim$9-10 kpc in the centre and 13-14 kpc at the edge of the disk, which is comparable to the extent of the extra planar emission seen in the edge-on spiral NGC 891 \citep{oos07}.  Looking at Fig. 5, the HI appears to surround the hot gas component in the outer parts.

\begin{figure}
\centering
\includegraphics[width=\columnwidth]{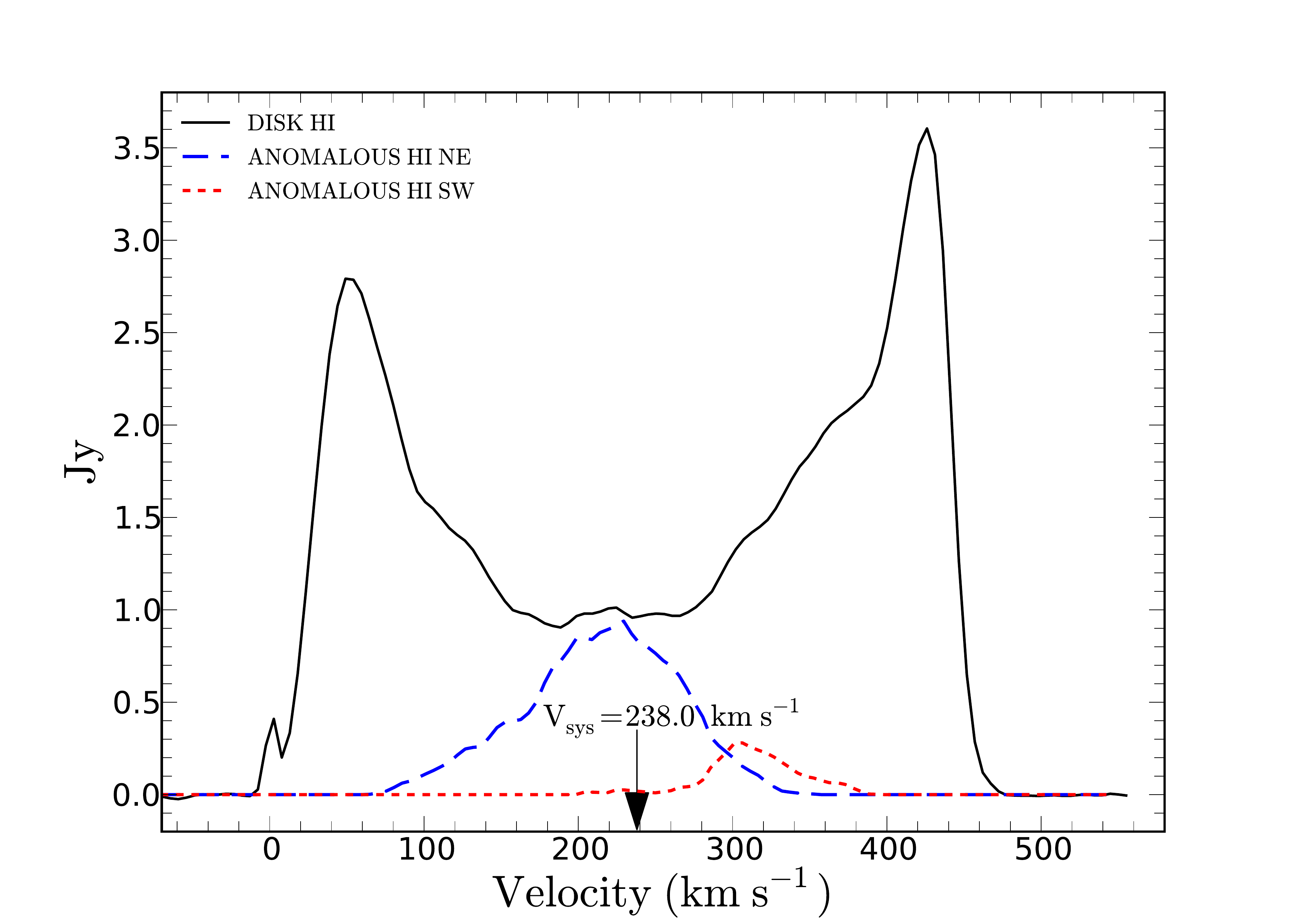}
\caption{Global profile (black) for the KAT-7 \hi\ observations with the kinematical V$_{\rm sys}$ indicated (see Sec.~\ref{sec:VFRC}).  Also plotted are the \hi\ profiles for the receding (red dotted) and approaching (blue dashed) components of the Halo (see Sec.~\ref{sec:expl}).  The channel width is 5 \kms\ .}
\label{fig:glprof}
\end{figure}

The global \hi\ profile of NGC 253 is given in Figure \ref{fig:glprof}. This was obtained using the task BLSUM in AIPS after primary beam correction. 
The asymmetry is clear with more \hi\ on the receding (SW) than on the approaching side.
From it, a mid-point velocity of 242$\pm 4$ \kms\ (50\% level) is found. Because of the asymmetry, this value has to be preferred to the intensity weighted mean velocity as
indicative of the systemic velocity. This can be compared to 243$\pm 2$ \kms\ found by the HIPASS survey \citep{kor04}. However, for the kinematical analysis of 
Sec.~\ref{sec:VFRC}, we will use the kinematically determined value. The profile widths at the 20\% and 50\%
levels are $\Delta$V$_{20} = 439 \pm 4$ \kms\ and $\Delta$V$_{50} = 412 \pm 4$ \kms, slightly larger than the \cite{kor04} values. A total \hi\ flux of
$728 \pm 36$ Jy.\kms\ is found (after primary beam correction), which corresponds to an \hi\ mass of $2.1 \pm 0.1$ $\times 10^{9}$ M$_{\odot}$ for a M$_{\rm HI}$/L$_{\rm B} \simeq 0.1$ at our adopted distance of 3.5 Mpc. This is similar to the mass of $2.0 \times 10^{9}$ M$_{\odot}$ found by \citet{boo05}. Note that in the next section, we will separate the disk and the kinematically anomalous (halo) components. The mass of the anomalous gas 
is given in Table~\ref{hipar}. 

\begin{figure}
\includegraphics[width=\columnwidth]{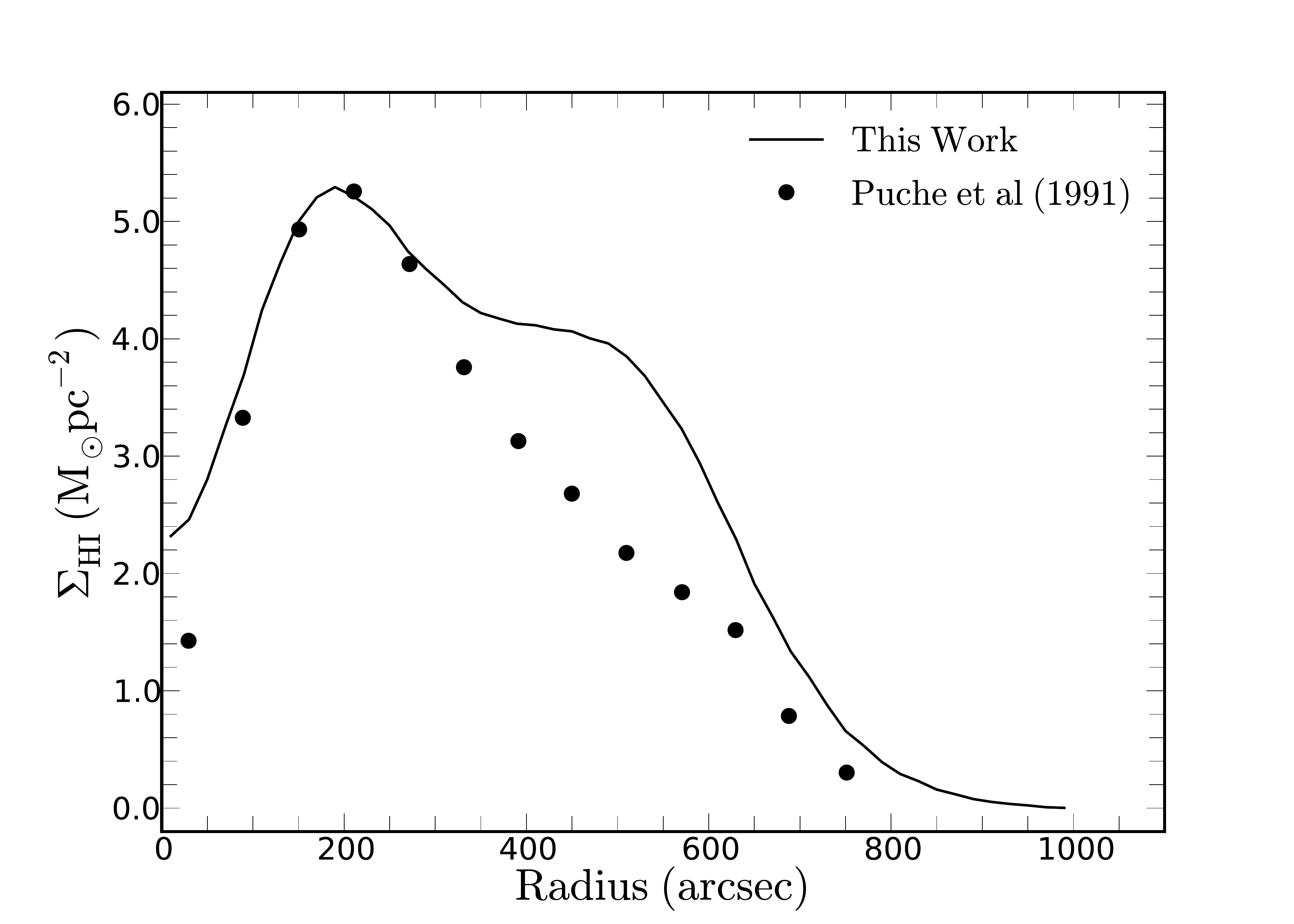}
\caption{Comparison of the VLA \protect\citep{puc91} and the KAT-7 \hi\ radial profiles.}
\label{fig:radprof}
\end{figure}

The radial \hi\ profile of NGC 253 is given in Figure \ref{fig:radprof} and compared to the profile derived by the previous VLA observations. It can be seen that, while the VLA was
detecting the bright \hi\ disk (down to 2.4 $\times$ 10$^{20}$ cm$^{-2}$), KAT-7 detects $\sim$33\% more flux (down to 1.3 $\times$ 10$^{19}$ cm$^{-2}$), mainly in the outer disk and in the halo. This is not only coming from the longer integration but also because KAT-7 can detect scales larger than 15\arcmin, scales invisible to the VLA because of the lack of short spacings. 

The derived \hi\ parameters are summarized in Table~\ref{hipar}. While the \hi\ disk of a late-type spiral is often much more extended than the stellar disk (D$_{25}$), this
is not the case for NGC 253. At a level of 1 M$_{\odot}$ pc$^{-2}$ ($1.25 \times 10^{20}$ cm$^{-2}$), the \hi\ disk of NGC 253 is equivalent to the stellar disk (isophotal major diameter, $D_{25}$) and even at a level of
10$^{19}$ cm$^{-2}$, it is only equivalent to the Holmberg diameter ($D_{HO}$: diameter at a surface brightness level of $\mu_B$ = 26.6 mag arcsec$^{-2}$). For example, NGC 300, another late-type spiral in the Sculptor group, has an
\hi\ disk 50\% larger than the optical disk \citep{puc90}.  What is exceptional with NGC 253 is not its radial extent but more its extent perpendicular to the disk.  Despite the apparent truncation of the \hi\ disk, NGC 253 is not \hi\ deficient with log(M$_{\rm HI}$) = 9.32, slightly larger than $<$log(M$_{\rm HI}$)$>$ for an isolated field Sc of 9.24 \citep{Sol96}.

\begin{table}
\centering
\caption{\hi\ parameters of NGC 253.}
\label{hipar}
\begin{tabular}{lc}
\hline\hline
Parameter  & Value\\          
\hline
M$_{\hi}$ (total)                                 	& $2.1 \pm 0.1$ $\times 10^{9}$ M$_{\odot}$  \\
M$_{\hi}$ (anom.)$^1$                               & $7.8 \times 10^{7}$ M$_{\odot}$   \\
M$_{\hi}$ (anom.)/M$_{\hi}$ (total)$^1$	& 3.5\%	 \\
$\Delta$V$_{20}$, (20\% level)		&439 $\pm$4 \kms   \\
$\Delta$V$_{50}$, (50\% level)		&412 $\pm$4 \kms   \\
V$_{\rm sys}$ (mid-point 50\%)          &242 $\pm$4 \kms   \\
V$_{\rm sys}$ (kinematical, TR)         &238 $\pm$4 \kms   \\
D$_{20}$, (diam. at 10$^{20}$ cm$^{-2}$ level)	&29 $\pm$2 kpc  \\
D$_{19}$, (diam. at 10$^{19}$ cm$^{-2}$ level)	&34 $\pm$2 kpc  \\
\hline
(1) These are lower limits.\\
\end{tabular}
\end{table}

\section{Separating the disk and halo $\hi$ Gas}
\label{sec:expl}
\begin{figure*}
\begin{minipage}{150mm}
\includegraphics[width=150mm]{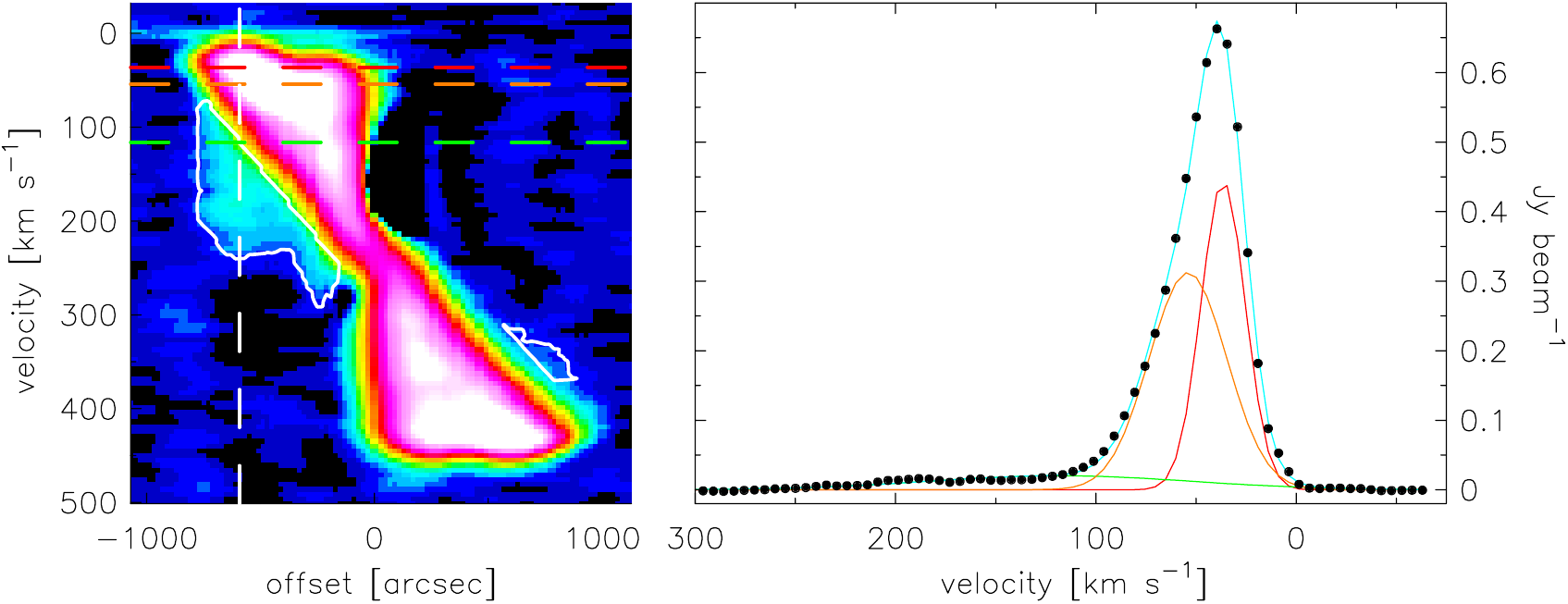}
\caption{(left) PV slice through the centre of the galaxy. The velocities corresponding to the peaks of the Gaussians have been indicated with the horizontal dashed lines, while the vertical dashed white line shows the location of the profile on the right.  (right) The red Gaussian traces the HI ridge, the orange Gaussian traces the beam-smeared emission, and the green Gaussian is fitting the anomalous HI.}
\label{fig:PV_AnHI}
\end{minipage}
\end{figure*}

The \hi\ total intensity map of NGC 253 clearly shows some of the \hi\ to be spatially separated from the main disk.  Furthermore, the kinematics of this spatially extended \hi\ component differ from that of the main disk.  
Position-Velocity (PV) slices, extracted parallel to the major axis of NGC 253, provide clear evidence of kinematically anomalous \hi.  Such a PV slice, passing through 
the centre of the galaxy, is shown in Figure \ref{fig:PV_AnHI}.  For the approaching (NE) half of the galaxy, it clearly highlights the presence of an anomalous \hi\ component 
spanning a range of radial velocities that is offset from the high-intensity emission of the main disk; toward the systemic velocity of the system.  Such anomalous \hi\ features are commonly seen in deep observations of nearby galaxies: e.g. NGC 891 \citep{oos07} and NGC 2403 \citep{fra01}. 

In order to isolate the anomalous \hi\ emission in NGC 253 from the regular \hi, PV slices aligned with the major axis were extracted along the full extent of the galaxy's minor axis. Each PV slice was visually inspected.  Any emission deemed to be kinematically anomalous was isolated, removed, and used to construct a new cube. In Figure \ref{fig:PV_AnHI}, the anomalous \hi\ is delimited by the white contours. This position-velocity diagram is representative of the majority of PV slices with anomalous \hi. 

\begin{figure*}
\begin{minipage}{150mm}
\includegraphics[width=\columnwidth]{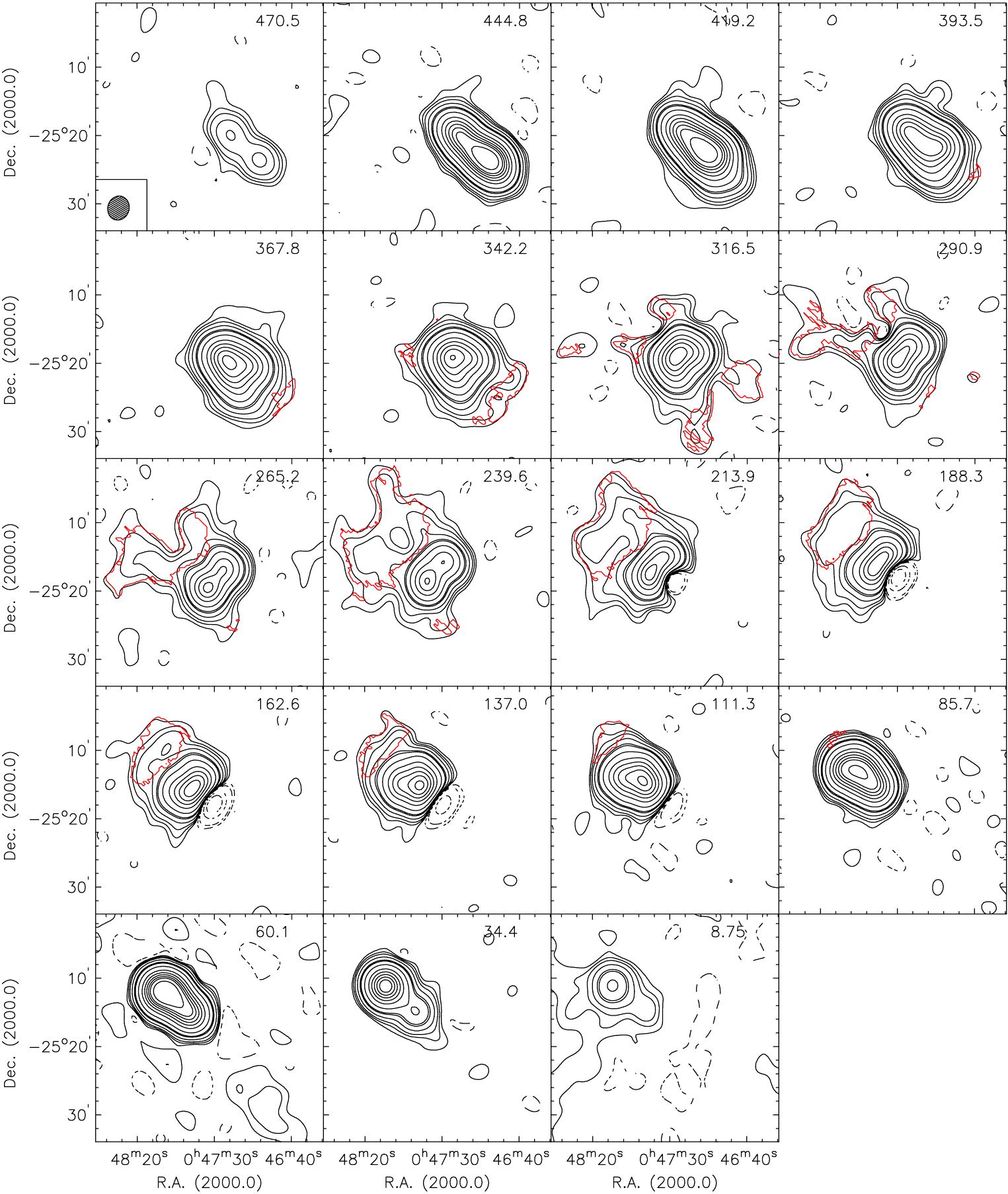}
\caption{Individual channel maps identifying the anomalous HI emission.The solid black contours are at levels of 0.8 0.6 0.5 0.4 0.3 0.2 0.1 0.05 0.04 0.02 0.01 0.006 0.003 Jy/beam.  The dashed black contours are at -0.1 -0.05 -0.01 -0.003 Jy/beam.  The red contour is at 0.003 Jy/beam. The beam is shown in the bottom left corner of the first channel.}
\label{fig:chmap}
\end{minipage}
\end{figure*}

\begin{figure*}
\centering
\includegraphics[width=150mm]{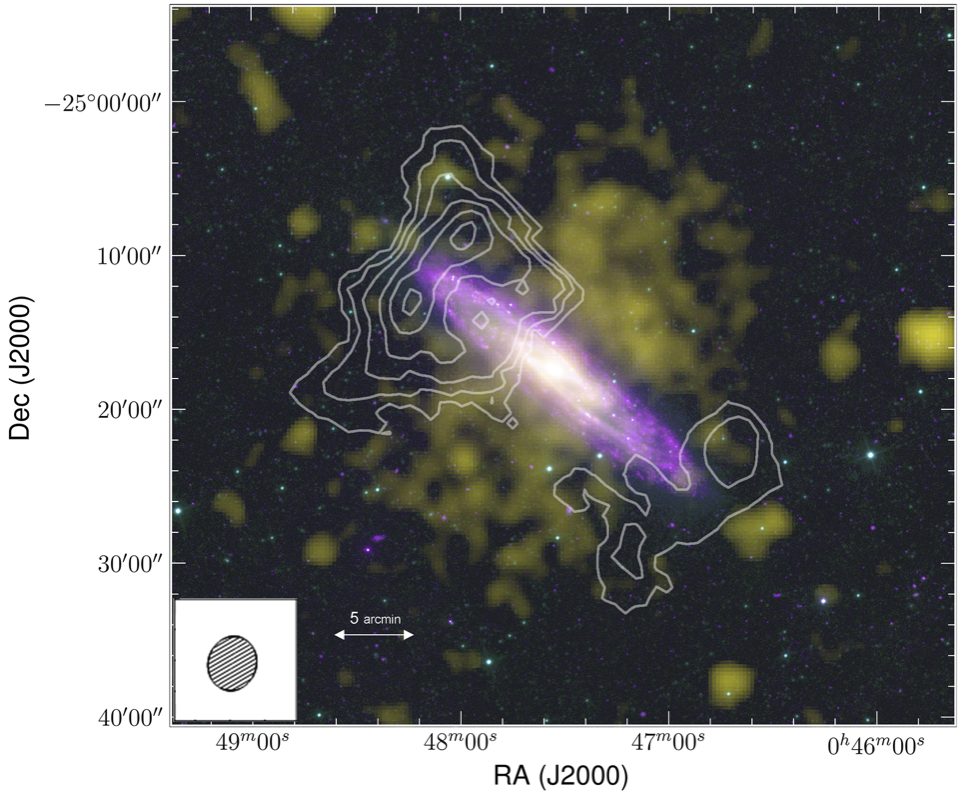}
\caption{Combined WISE (RGB) + FUV (magenta) + X-rays (yellow: Rosat) + HI contours of the anomalous \hi\ (scale: 1\arcmin $\sim$ 1 kpc). The contours correspond to 0.03, 0.2, 0.5, 1.0, 1.5 2.0 Jy/beam  \kms.}
\label{fig:WUVXhaloHI}
\end{figure*}

The spatial distribution of the anomalous \hi\ in that new cube is delimited by red contours in the channel maps shown in Figure \ref{fig:chmap}.  Summing all the gas gives an extended component on the approaching half of the galaxy and a smaller component on the receding side as can be seen in Figure \ref{fig:WUVXhaloHI}. 
The mass of that component is estimated to be $\sim$3.5\% of the total \hi\ mass of the galaxy ($\sim7.8 \times 10^{7}$ M$_{\odot}$). This is equivalent to the mass of the  anomalous gas found by \citet{boo05}. 

\begin{figure*}
\centering
\begin{tabular}{ccc}
\includegraphics[width=\columnwidth]{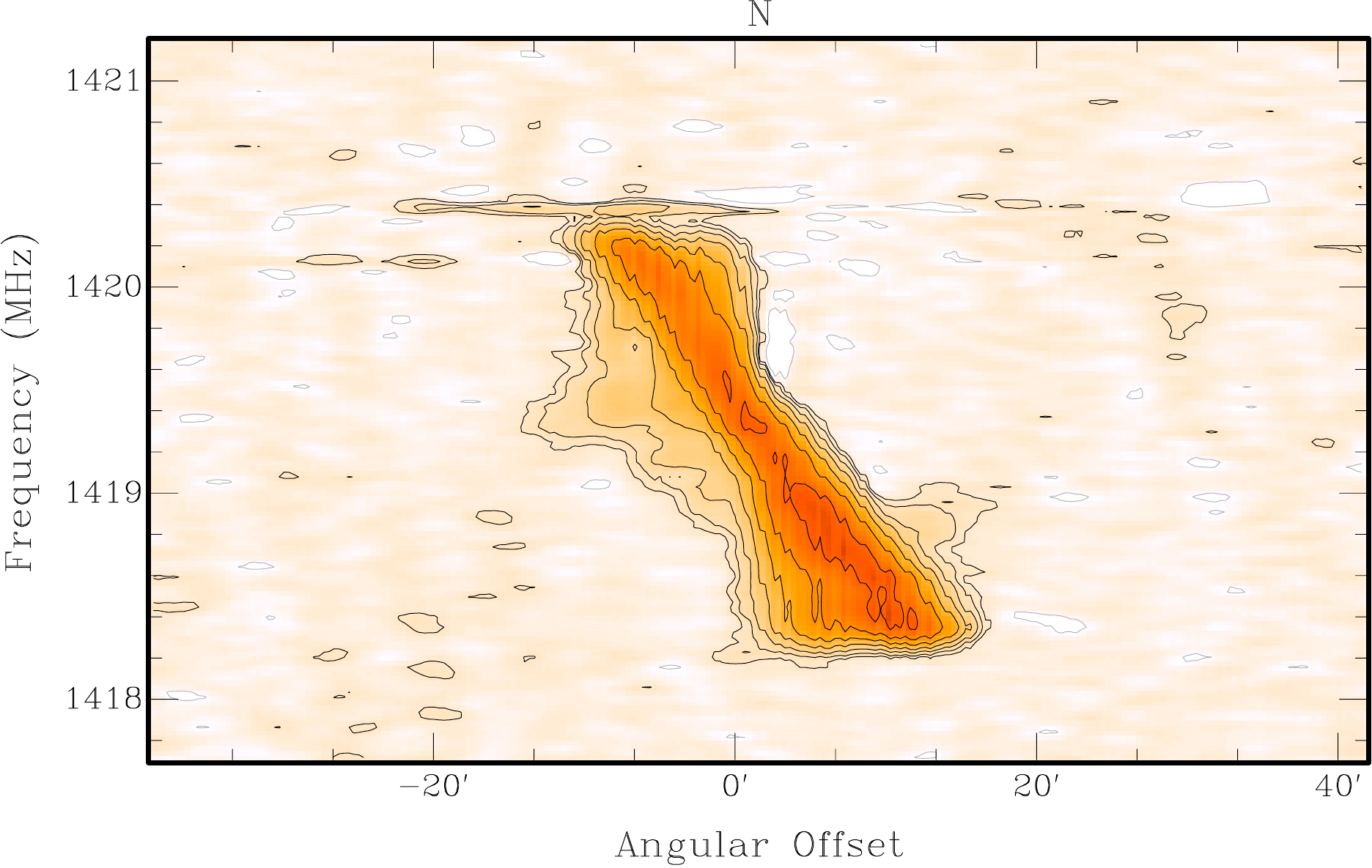}\\
\includegraphics[width=\columnwidth]{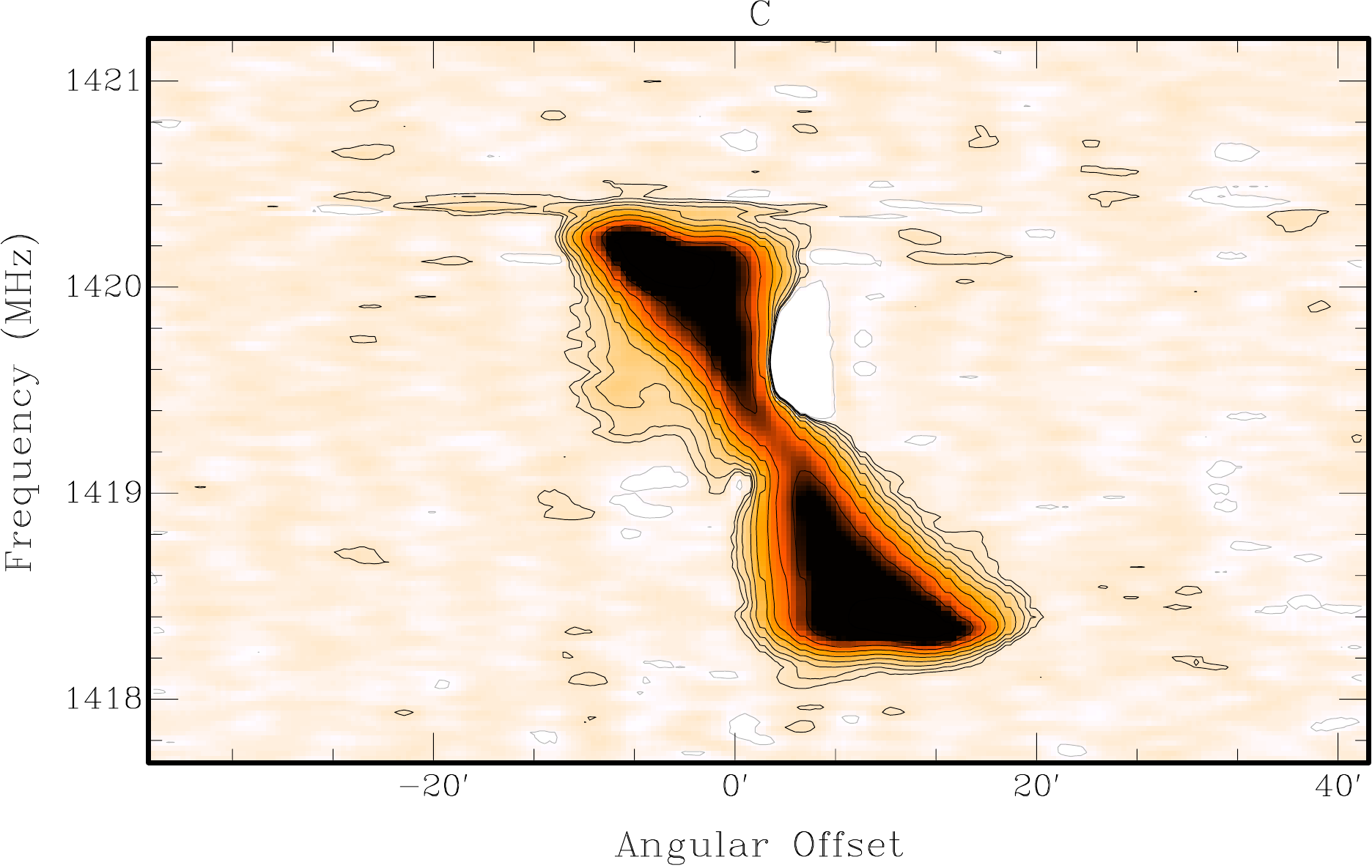}\\
\includegraphics[width=\columnwidth]{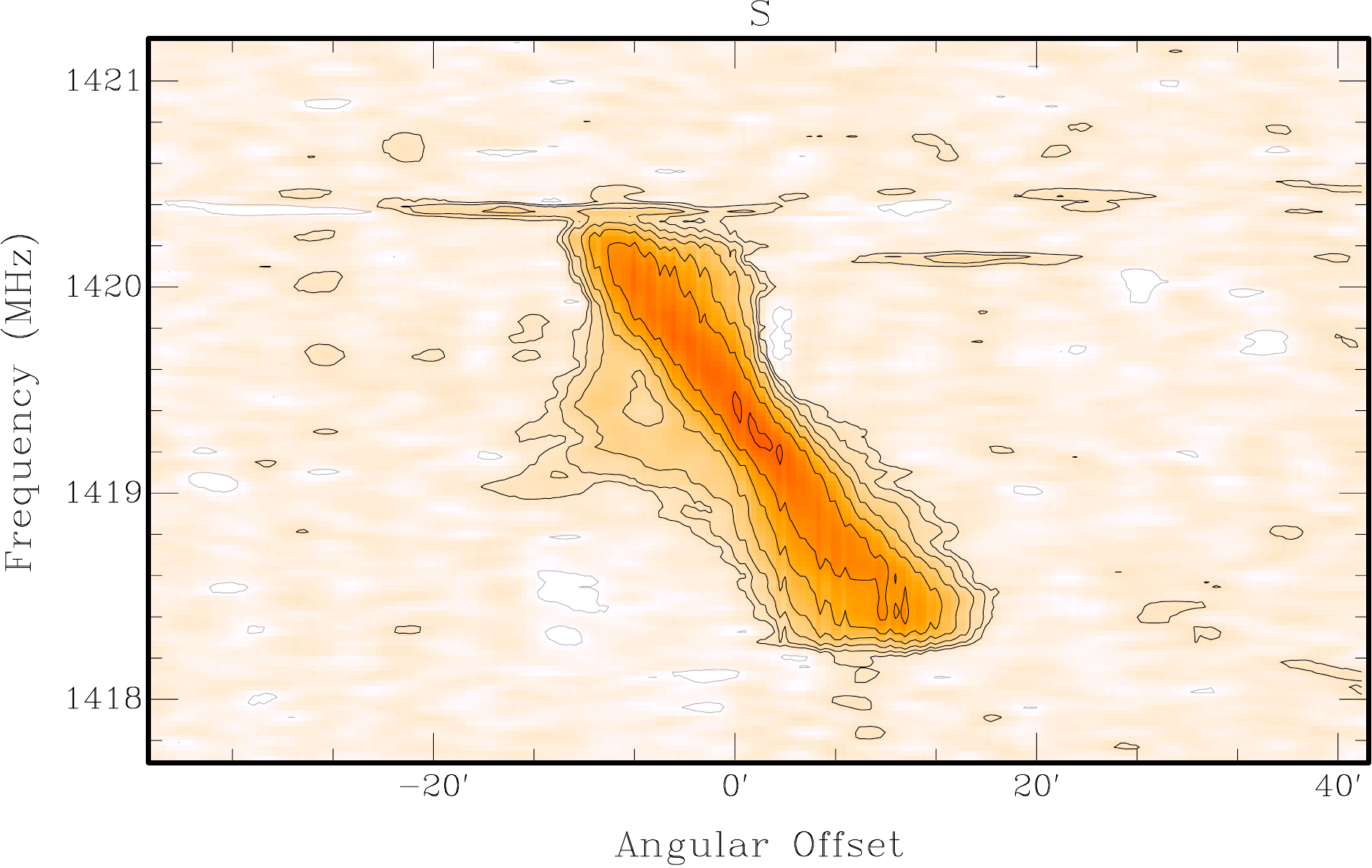}\\
\end{tabular}
\caption{PV slices parallel to the major axis. One is on the centre while the other two are 4\arcmin\ NW and SE of the major axis along the minor axis. Contour levels are -3, 3, 6, 12, 24, 48, 96 mJy/beam}
\label{fig:PVslices}
\end{figure*}

Naturally, this should be seen as a lower limit since our
technique has most probably missed a large fraction of the "anomalous" gas, especially along the minor axis to the NW and SE due to projection effects. This can be seen on the PV slices 
along the minor axis shown in Figure \ref{fig:PVslices}. However, this will not be a problem for the kinematical study since, as we will see in Sec. \ref{sec:VFRC}, the data along the minor axis will be excluded.

In order to generate a velocity field representative of the circular kinematics of the \hi\ disk, we again used PV slices aligned with the major axis of the galaxy, and fitted interactively three Gaussians to all of the line profiles making up a slice. Great care was taken in selecting the Gaussian associated with the disk of the galaxy.  This was done by simultaneously viewing a position velocity slice and the fitted Gaussians. The line profile along the white dashed line on the left part of Figure \ref{fig:PV_AnHI} is shown on the right.  For this particular line profile, the red Gaussian parameterizes the radial velocity of the high-intensity HI ridge in the PV slice.  This is the component of the line profile  associated with the regularly-rotating \hi\ disk of the galaxy.  The orange Gaussian, with its peak shifted toward systemic velocity, represents the beam-smeared component of the line profile.  The green Gaussian captures the kinematically anomalous \hi\ emission. 

\begin{figure}
\centering
\includegraphics[width=\columnwidth]{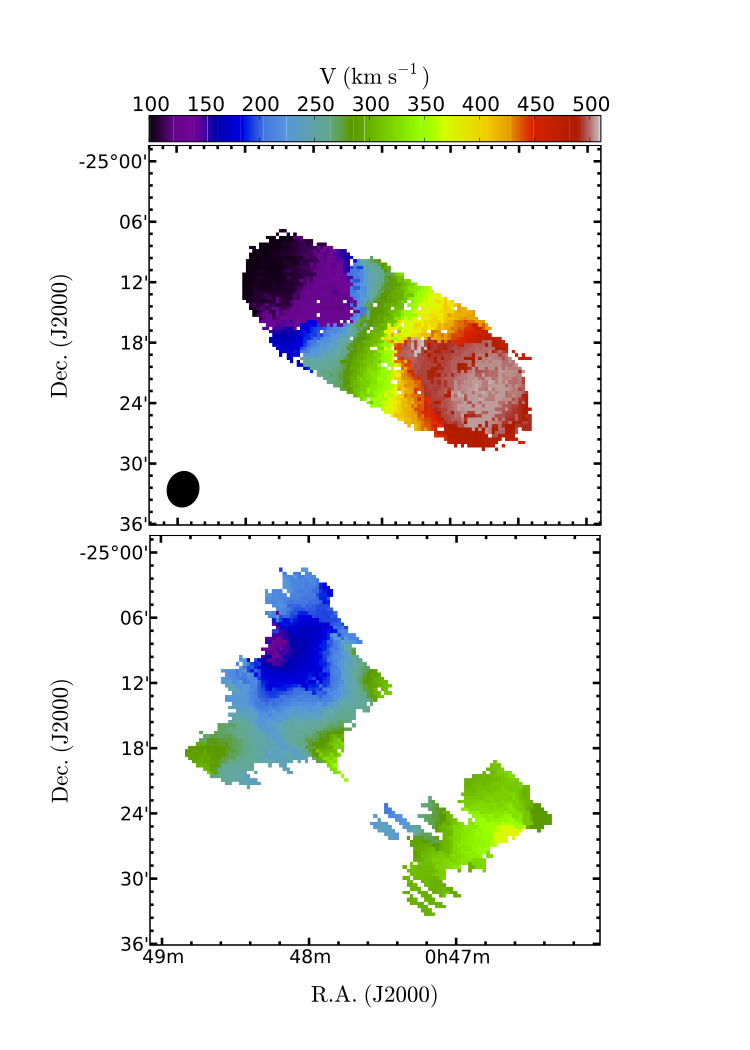}
\caption{Velocity fields of the disk (top) and anomalous (bottom) \hi.}
\label{fig:VELFI}
\end{figure}

Stepping along the minor axis of the galaxy, this procedure was repeated for all the PV slices aligned parallel to the major axis.  In doing so, a velocity field representing the circular kinematics of the HI disk of NGC 253 was built up pixel-by-pixel.  When none of the Gaussians fitted to a line profile provided a reliable measure of the circular rotation, a blank was assigned to the corresponding position in the velocity field. We consider the velocity field shown in Figure \ref{fig:VELFI} (top panel) to better represent the circular kinematics of NGC 253 than the traditional intensity-weighted-mean velocity field.  From the cube built for the anomalous \hi, a velocity field was also obtained by a moment analysis and is shown in the bottom panel of Figure \ref{fig:VELFI}. 
\section{\hi\ Kinematics of the disk and halo gas}
\label{sec:VFRC}
To derive the RCs, we used the implementation of the tilted ring model \citep{rog74} in the GIPSY  \citep{vdh92} task {\sc ROTCUR} \citep{beg89, beg87}.
In the tilted ring model, a set of concentric rings is used to describe the motion of the gas in the galaxy. The gas is assumed to be in circular motion. Each ring is characterized by 
a set of 5 orientation parameters, namely: a rotation centre $(x_c,y_c)$, a systemic velocity $V_{sys}$, an inclination $i$, a Position Angle $PA$ and by
a rotation velocity $V_{C}$. Naturally, the rotation centre $(x_c,y_c)$ and the systemic velocity $V_{sys}$ should be the same for all the rings, at least within the optical disk,
but $i$ and $PA$ will vary if the \hi\ disk is warped outside D$_{25}$.

The line of sight velocity at any $(x,y)$ position in a ring with radius $R$ is given by

\begin{equation}
\label{ }
V(x,y) = V_{sys} + V_{C} sin(i) cos(\theta) 
\end{equation}
 where $\theta$ is the position angle with respect to the receding major axis measured in the plane of the galaxy. $\theta$ is related to the 
 actual $PA$ in the plane of the sky by
 \begin{equation}
\label{ }
cos(\theta) = \frac{-(x  - x_{0}) sin(PA) + (y - y_{0}) cos(PA)}{R} \\\\
\end{equation}
\begin{equation}
\label{ }
sin(\theta) = \frac{-(x  - x_{0}) cos(PA) + (y - y_{0}) cos(PA)}{R cos(i)}
\end{equation}
\subsection{Tilted-ring model for the \hi\ disk}
A $ |cos{\theta}|^2$ weighting function and an exclusion angle of 60$^{\rm o}$ about the minor axis have been used to give maximum weight 
to the velocity points close to the major axis and to minimize the influence of large deprojection errors close to the minor axis 
in view of the large inclination of the galaxy.  We used two rings per beam size.

The method consists of finding for each ring the best set of the 5 orientation
parameters $(x_c,y_c)$,  $V_{sys}$, $i$ and $PA$ 
which minimizes the dispersion of $V_C$ inside the ring.
The following procedure is used:
\begin{itemize}
\item The rotation center $(x_c,y_c)$ and the systemic velocity $V_{sys}$ are determined first
by keeping $i$ and $PA$ fixed (using the optical values). The rotation center and
the systemic velocity have to be determined simultaneously since they are correlated. 
\item Now, keeping $(x_c,y_c)$ and $V_{sys}$ fixed, $i$ and $PA$ are fitted to map any possible
warp of the \hi\  disk. The warps
usually start just outside the optical. However, since the \hi\ disk in NGC 253 is about the same size as the optical,
we may not reach the warp region. Here also, $i$ and $PA$ have to be determined
simultaneously since they are correlated.
\item The previous two steps were done using the data of both sides of the galaxy together. Using the same fixed
$(x_c,y_c)$ and $V_{sys}$, the previous step is repeated for the approaching and
receding sides separately to identify possible departures from axisymmetry.
\end{itemize}

\begin{figure}
\includegraphics[width=\columnwidth]{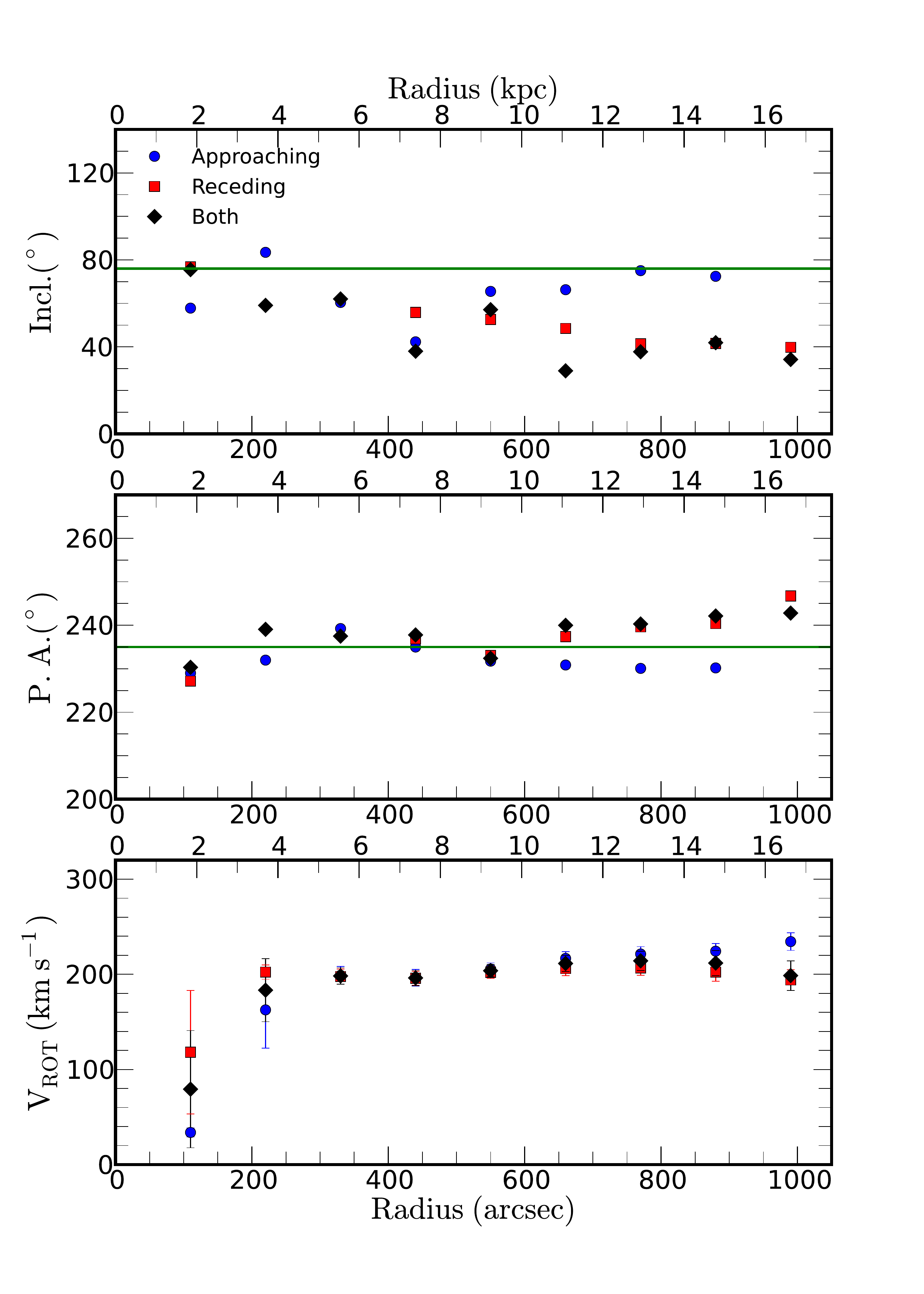}
\caption{ROTCUR solution and derived RC for the \hi\ disk.The green lines show our adopted $PA = 235^o$ and $i = 76^o$.}
\label{fig:disk_rocur}
\end{figure}

The ROTCUR solutions (approaching, receding, both sides)  are shown in Figure \ref{fig:disk_rocur} with $PA$ and $i$ free to vary.  Because of the low resolution of our data, it can be seen that, while $PA$ is fairly well determined ($<PA> = 235^o \pm 5^o$, $i$ is not constrained at all. It is clear, looking at the galaxy, that inclinations $\sim 40^o$, as seen for $r > 10$\arcmin, are completely unrealistic.  However, much higher resolution data (30\arcsec) exist from Fabry-Perot observations \citep{hla11}.  In their Figure 10, we can see that the inclination is nearly constant at $i = 76^{\rm o} \pm 4^{\rm o}$.  Additionally, similar values of i ($78.5^{\rm o}$) and PA ($232^{\rm o}$) have been inferred from CO observations \citep[e.g.][]{sof99, hou97}.  In the case of NGC 253, adopting constant $PA$ and $i$ is reasonable since, as we have seen in Sec.~\ref{sec:HICD}, the \hi\ disk has the same size as the optical disk and warps usually start past the optical radius.  We thus adopt constant values of $PA = 235^o$ and $i = 76^o$ (green lines in Fig. \ref{fig:disk_rocur}) to derive our adopted RC. 

A similar approach was used by \citet{puc91} using higher resolution \hi\ data. The adopted RC for the \hi\ disk (solution for both sides with $PA$ and $i$ fixed) 
is given in Table \ref{n253RC}. An important point to notice is the large errors on the first two points in the inner parts. This is expected for at least three reasons: 1- the strong absorption feature in the centre; 2- the small quantity of \hi\ in the centre (cf. Fig, \ref{fig:radprof}); 3- the strong disturbance expected from the starburst in the inner regions.
\subsubsection{Three-dimensional  modelling}
\label{sec:3D}
As a further test on our choice of $i$, three different tilted ring models were generated, with constant inclinations of 66, 76 and 86 degrees.  For each of the models, all the other parameters ($PA, V_{sys}, X_c, Y_c$) were kept fixed to the same constant values, allowing only $V_c$ to vary with radius.  The tilted ring models were then used to generate 3D models of the NGC 253 \hi\ data cube. This was done using the task GALMOD in GIPSY.  For the radial distribution of \hi\, we adopted the \hi\ surface density profile shown in Figure \ref{fig:radprof}.  A fiducial \hi\ disk scale height of 0.5 kpc was used for each model.  The resulting model cubes were smoothed to the spatial resolution of the data, and the fluxes scaled based on the NGC 253's integrated flux density of 728 Jy km/s.  

\begin{figure}
\includegraphics[width=\columnwidth]{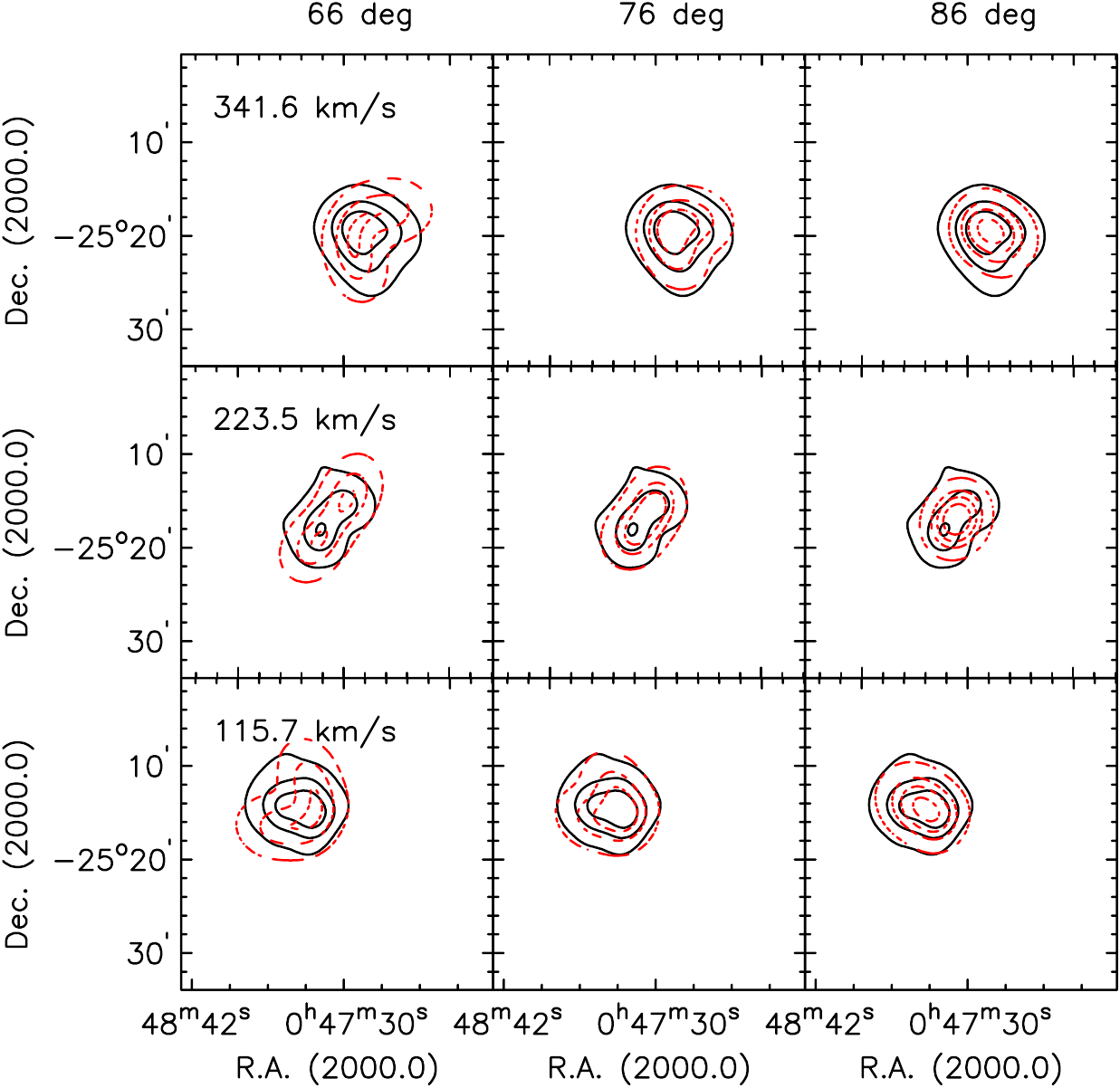}
\caption{Comparison of the data (black) with 3D-models (red-dashed) using ROTCUR solutions with $i$ = 66$^o$ (left), 76$^o$ (centre) 
and 86$^o$ (right) for 3 different channels. The contours are at levels of 0.03, 0.2, 0.4, 0.7 Jy/beam.}
\label{fig:3Dmod}
\end{figure}

These models, each based on a different inclination, were then compared to the data on a channel-by-channel basis.  Doing so clearly showed that the $ i = 76^o$ model consistently matches the data better than the models with $i =66^o$ and $i = 86^o$.  For three different channels, Figure \ref{fig:3Dmod} shows the comparisons between the models and the data.  Clearly, $i = 66^o$ (left) produces emission too elongated in the minor axis direction and $i = 86^o$ (right) too narrow. Based on these results, we adopt $i = 76^o$ as the preferred inclination for the \hi\ disk of NGC 253.  This inclination, together with the rotation curve from the tilted ring model, accurately describes the kinematics of the HI disk.  
\subsubsection{Comparison of the RC with the data}
\begin{figure}
\centering
\begin{tabular}{cc}
\includegraphics[width=\columnwidth]{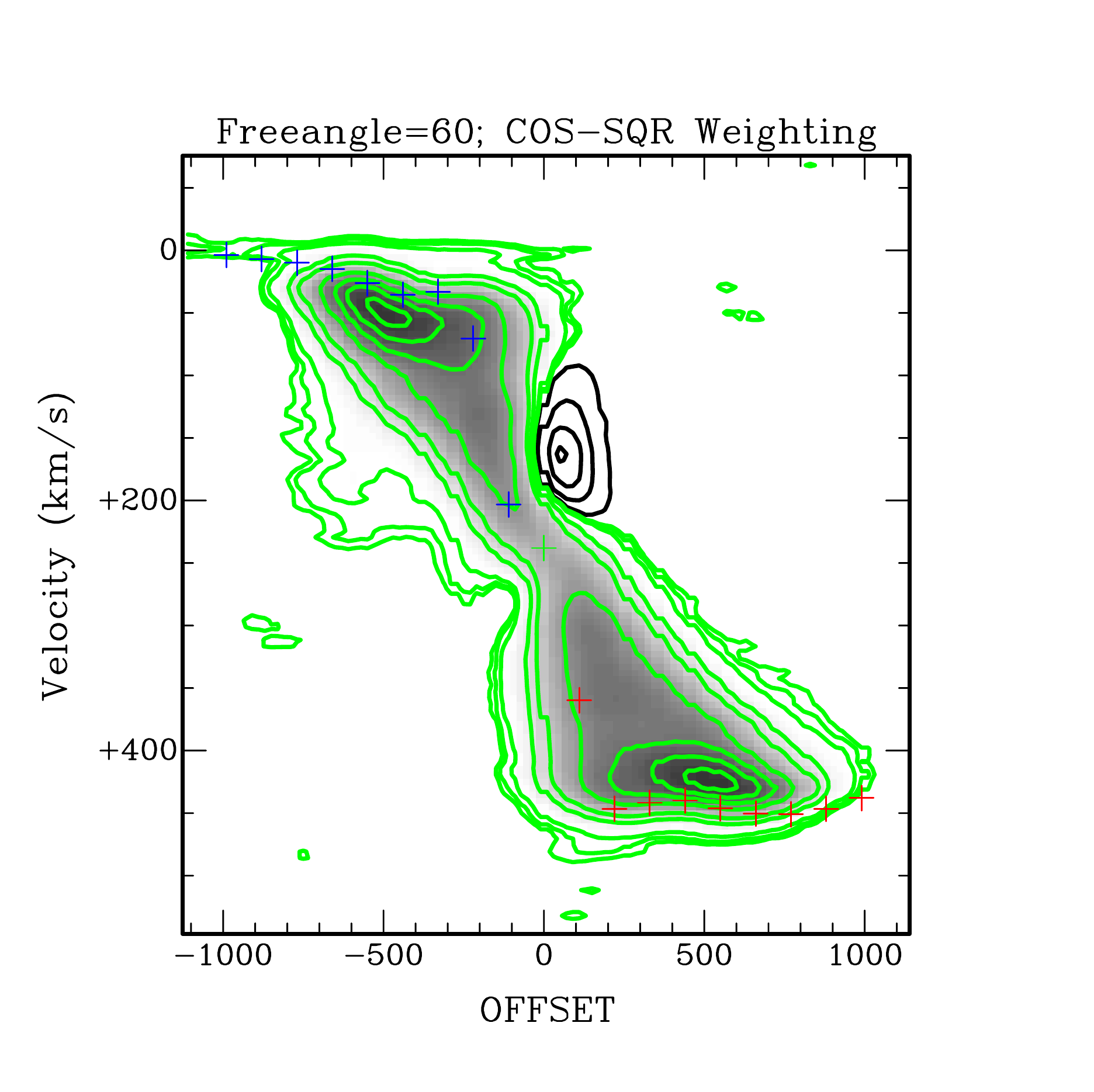}\\
\includegraphics[width=\columnwidth]{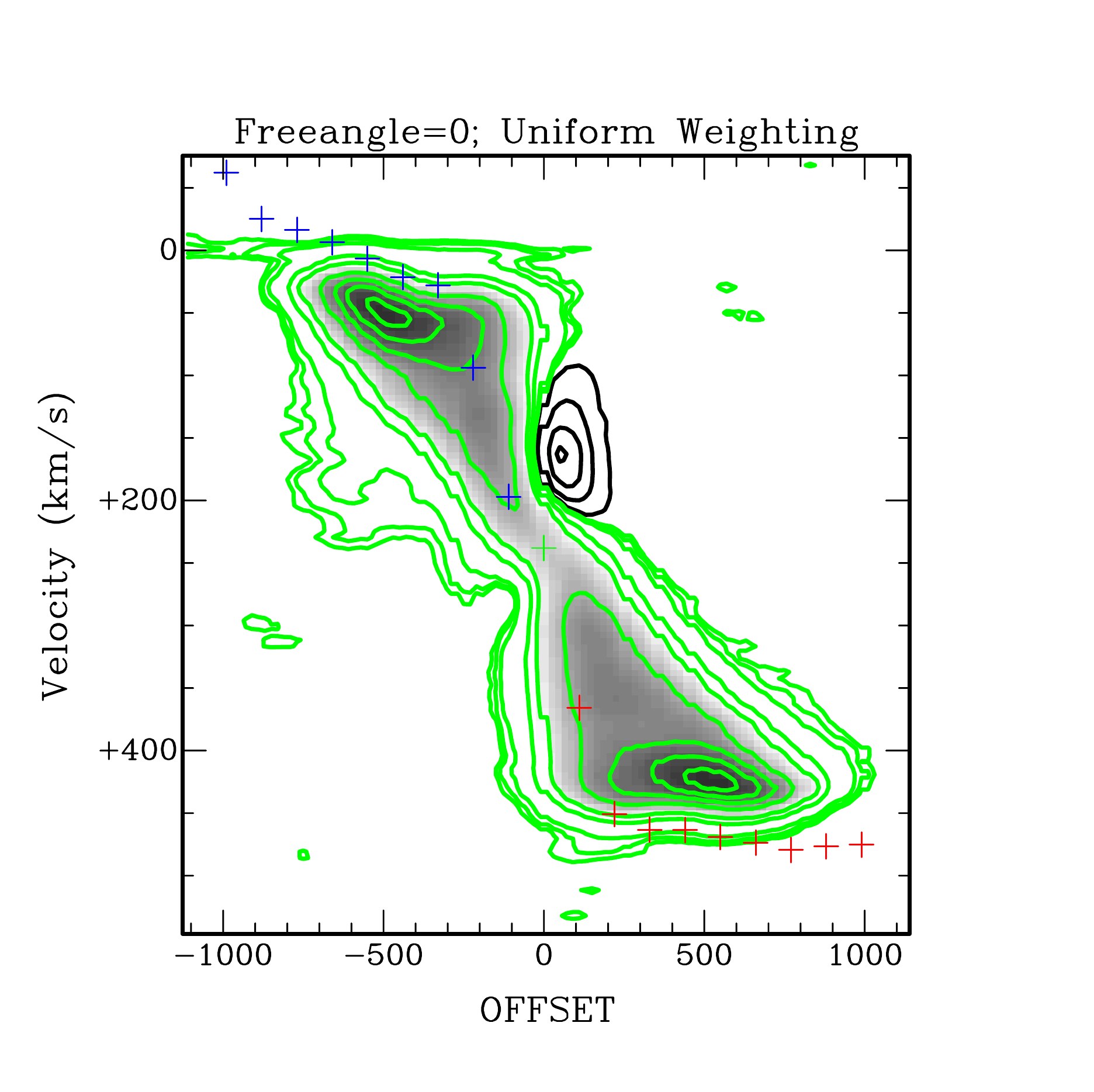}\\
\end{tabular}
\caption{PV diagrams along the major axis with the receding (red pluses) and approaching (blue pluses) RC superimposed
using the adopted kinematical parameters with a free angle of $60^o$ and a $ |cos{\theta}|^2$  weighing function at the top and no free angle  and no weighing function at the bottom. 
The black contours are negative contours around the absorption feature.}
\label{fig:pv}
\end{figure}

Finally, the best test to check that our RC accurately represents the kinematics of the disk is to look at the superposition of
the receding and the approaching sides on a Position-Velocity (PV) diagram obtained along the major axis, as shown in Figure \ref{fig:pv} (top panel). It can be seen that the deprojected velocities follow very closely the ridge of the emission, which shows that our RC is a good representation of the kinematics on the major axis.
Note on the PV diagram the anomalous \hi\ around 200 \kms. We should point out the presence of Galactic \hi, which can be seen around 0 \kms. Even though we did exclude the channels with obvious Galactic emission, it can be seen that some of it is still present. However, since there is much more \hi\ on the receding side than on the approaching side, very
few data points from the approaching side contribute to the last points of the solution for both sides (cf. Fig. \ref{fig:disk_rocur}) and make us confident in our adopted RC.

One last point needs to be emphasized. At the beginning of the section, we mentioned that a $ |cos{\theta}|^2$ weighting function and an exclusion angle 
of 60$^{\rm o}$ about the minor axis have been used to minimize the influence of large deprojection errors close to the minor axis 
in view of the large inclination of the galaxy. Typically, when a galaxy is not highly inclined (p.e. $i = 30-45^o$), all the data can be used. For intermediate inclinations (p.e. $i = 45-60^o$), a free angle of $\sim 30^o$ and a $ |cos{\theta}|$ weighting function should be used to give more weight to the data points close to the major axis.
However, for high inclinations (p.e. $i > 75^o$), it is suggested to use a larger exclusion angle ($\sim 60^o$) and a $ |cos{\theta}|^2$  weighting function. 

To illustrate this, we ran a ROTCUR model for NGC 253 with no exclusion angle and no weighting function. The resulting RCs for the approaching and receding sides are overlaid on the PV-diagram of the major axis in the bottom panel of Figure \ref{fig:pv}. We see that this solution is inconsistent with the kinematics observed on the major axis and must be discarded. This illustrates well that one has to be very careful when deriving the kinematics of highly inclined galaxies, especially with ROTCUR.

\begin{table}
\centering
\caption{Rotation velocities and errors for the \hi\ disk of NGC 253.}
\label{n253RC}
\begin{tabular}{ccc}
\hline\hline
Radius    &   V$_{\rm rot}$   &  $\Delta{\rm V}$ \\
(arcsec)  & (\kms)               & (\kms)       \\
\hline
110    &       79.3 & 61.6  \\
220    &   183.3 & 33.1  \\
330    &   198.3 &  8.6  \\
440    &   196.2 &  8.0  \\
550    &    203.7 &  6.4  \\
660    &    211.4 &  8.7  \\
770    &   214.1 & 10.1  \\
880    &   211.8 &  13.2 \\
990    &  198.6 &  15.5 \\
1100	  & 186.7 & 16.5 \\
\hline
\end{tabular}
\end{table}

\begin{figure}
\includegraphics[width=\columnwidth]{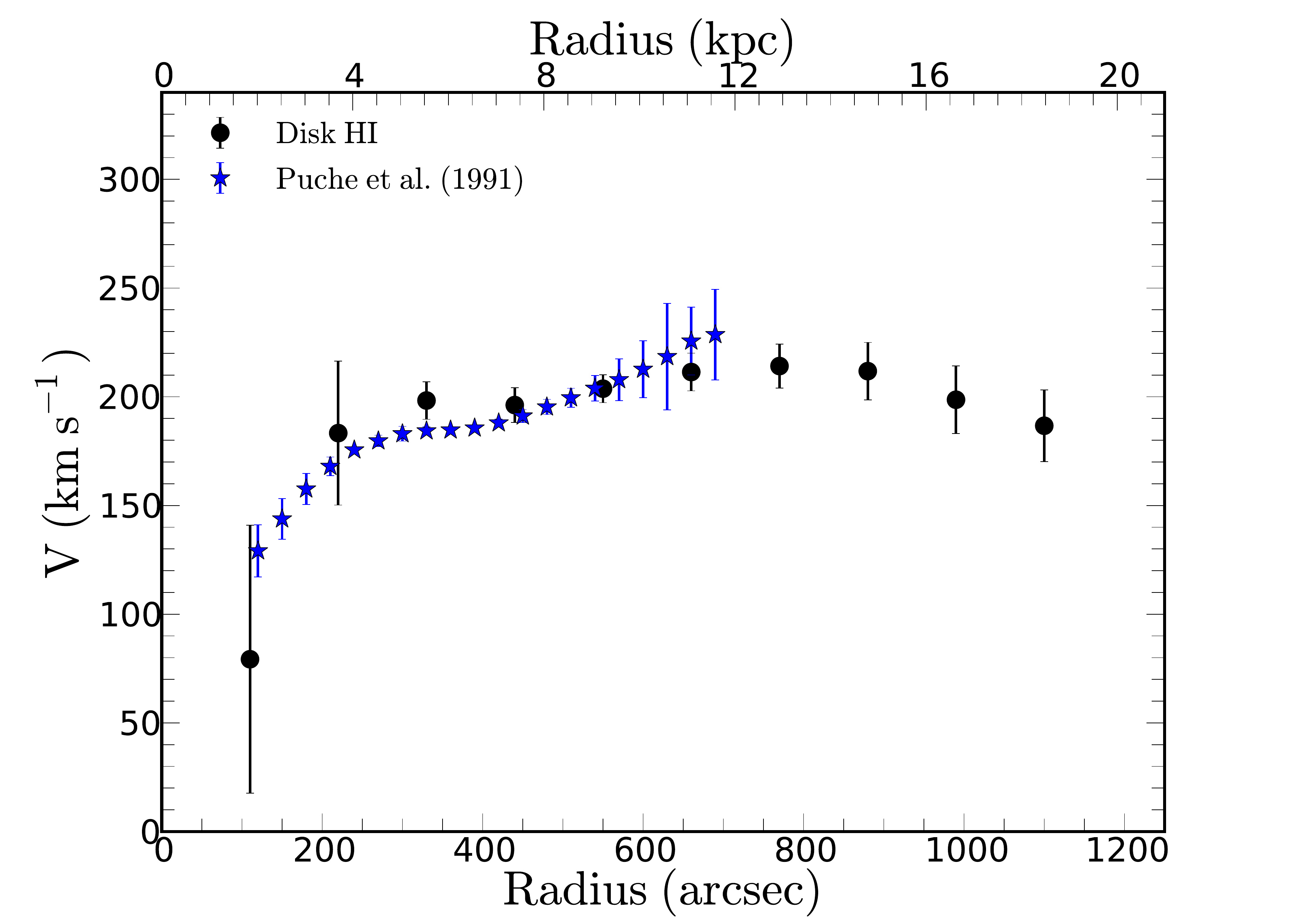}
\caption{Comparison of the VLA \& KAT-7 rotation curves.}
\label{fig:RC_VLA_KAT7}
\end{figure}

Figure \ref{fig:RC_VLA_KAT7} compares our adopted RC for the disk with the one derived by \citet{puc91} using
VLA observations. We see that out to r $\sim 12$ kpc (the last measured velocity point with the VLA data), 
both rotation curves agree well within the errors, with the exception of our velocity at 5.5\arcmin, which is $\sim 15$ \kms\ larger.
As for the KAT-7 RC, it rises to a velocity of $\sim 200$ \kms at 5.5\arcmin, stays more or less flat out to $\sim 15\arcmin$
and than declines by $\sim 20$ \kms out to the last point. We will discuss this possible decline in Sec.~\ref{sec:decl}.

\subsection{Tilted-ring model for the anomalous \hi}

\begin{figure}
\includegraphics[width=\columnwidth]{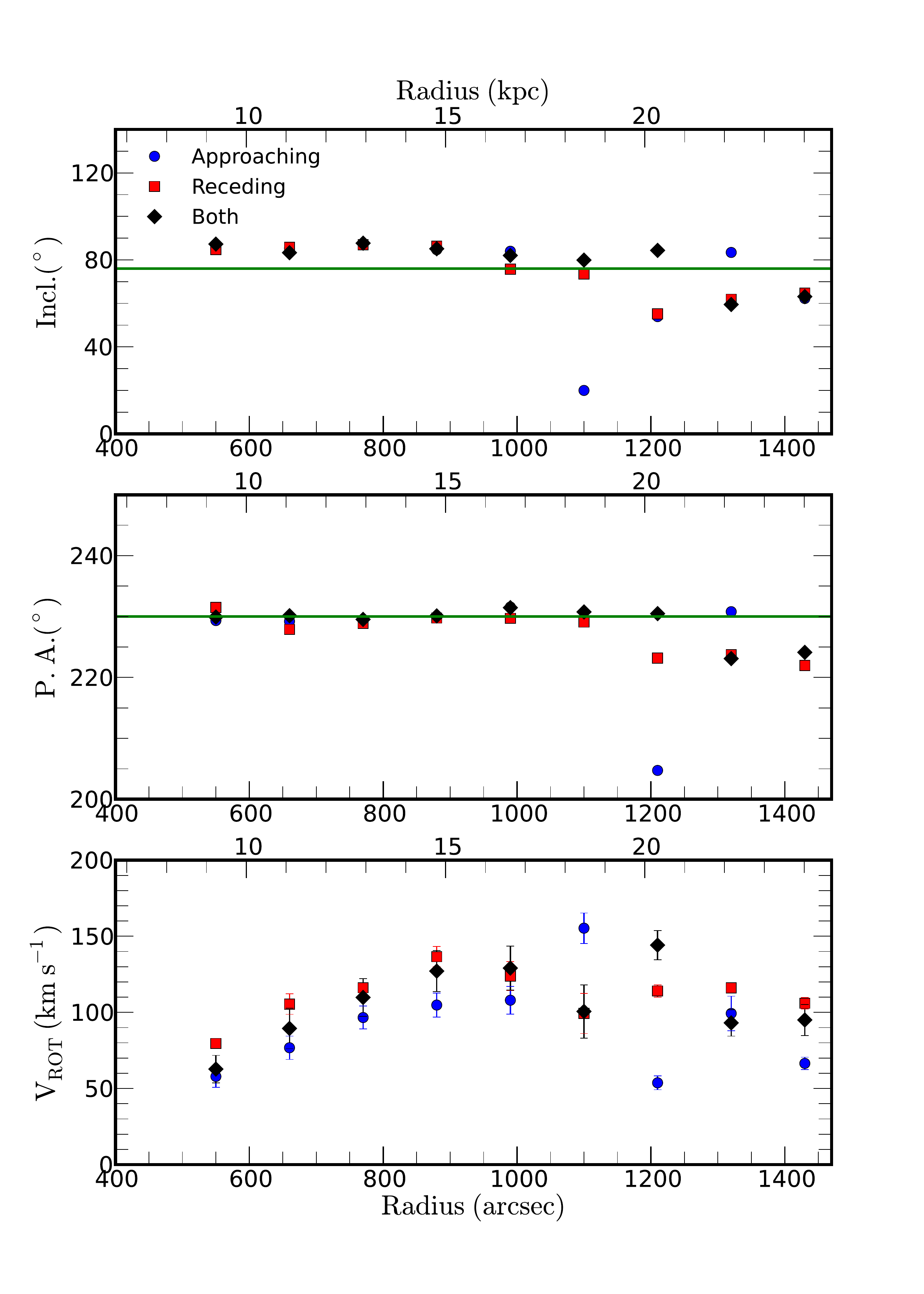}
\caption{ROTCUR solution and derived RC for the anomalous \hi. The green lines show our adopted $PA = 230^o$ and $i = 76^o$.}
\label{fig:ano_rocur}
\end{figure}

\begin{figure}
\includegraphics[width=\columnwidth]{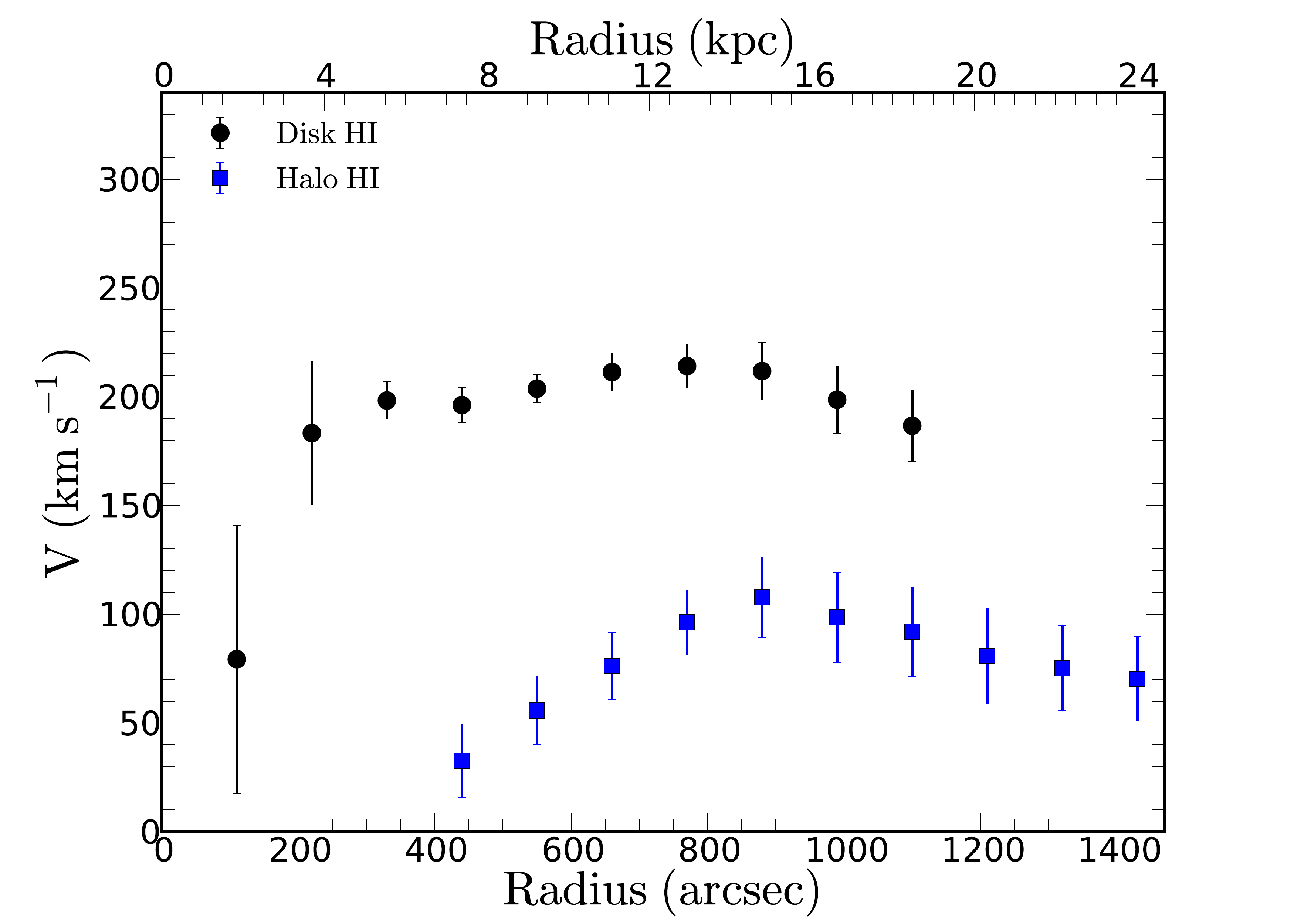}
\caption{Comparison of the disk kinematics and of the anomalous \hi\ kinematics.}
\label{fig:C_disk_ano}
\end{figure}

The kinematical solution for the halo gas is shown in Figure \ref {fig:ano_rocur}.
Tilted ring models were fitted to the velocity field of the anomalous HI (shown at the bottom of Fig. \ref{fig:VELFI}). When allowed to freely vary with radius, 
it is seen that $PA$ and $i$ are much better constrained for the anomalous component than for the disk.
Despite the small change of $PA$ for the last two points, 
it was decided to adopt constant $PA$ and $i$ (green lines).
The rotation velocities for the anomalous gas are given in Table \ref{n253anomalous} 
and are shown in blue in Figure \ref{fig:C_disk_ano}, along with the rotation velocities for the disk. 
It is different from the RC seen in the bottom panel of Fig \ref{fig:ano_rocur} because 
$PA$ and $i$ were kept fixed (green lines).

\begin{table}
\centering
\caption{Rotation velocities and errors for the anomalous \hi\ of NGC 253.}
\label{n253anomalous}
\begin{tabular}{ccc}
\hline\hline
Radius    &   V$_{\rm rot}$   &  $\Delta{\rm V}$ \\
(arcsec)  & (\kms)               & (\kms)       \\
\hline
440    &   32.6 &  16.9  \\
550    &    55.7 &  15.8  \\
660    &    76.1 &  15.4  \\
770    &   96.3 & 15.0  \\
880    &   107.8 &  18.5 \\
990    &  98.6 &  20.8 \\
1100	  & 91.9 & 20.7 \\
1210	  & 80.7 & 22.1 \\
1320	  & 75.2 & 19.5 \\
1430	  & 70.2 & 19.4 \\
\hline
\end{tabular}
\end{table}

This RC for the anomalous gas is surprisingly very regular. 
It rises out to $\sim$15 kpc and then declines slowly out to the last measured velocity point.
In the outer parts, the anomalous gas is kinematically lagging the disk gas by $\sim 100$ \kms. Such lagging of extra-planar gas is also seen in other galaxies \citep[see e.g.][]{els14, hes09, hea07, fra01}. 

\section{Discussion}
\label{sec:dis}

\subsection{Is the Rotation Curve of NGC 253 really declining ?}
\label{sec:decl}

Since the discovery in the 1970s that the RC of most spiral galaxies remains flat out to their last measured point \citep{fre70, bos81}, the obvious question is: where do the disks of spiral galaxies end? Especially in the case of spiral galaxies, the \hi\ surface density declines with increasing radius to a point where it decreases abruptly and the \hi\ disk seems to be truncated. This truncation is usually around surface densities $\sim$$10^{18}$-$10^{19}$ cm$^{-2}$ (\hi\ self-shielding limit). 
This does not necessarily mean that there is no more hydrogen further out. As suggested by \citet{bfq97}, it may just be that the ambient radiation field out there is sufficiently strong that cold gas gets ionized \citep{mal90, sil76}. While the UV background radiation may not be sufficiently powerful to be responsible for this ionization, the situation may be different in a starburst object like NGC 253. 

We have seen in Sec. \ref{sec:HICD} that the \hi\ disk of NGC 253 only extends out to the optical radius and the WISE mid-IR extent. But if, like most late-type spirals, there is hydrogen further out and if, also as seen in most spirals, the cold gas disk is warped, it could be exposed to the intense UV radiation of the starburst nucleus and be ionized. This was the motivation for \citet{bfq97} and \citet{hla11} to try to detect H$\alpha$ emission in the outer parts of NGC 253 using Fabry-Perot (FP) interferometry, and perhaps succeed to find where the mass (luminous \& dark) ends. 

Both groups detected H$\alpha$ at the limit of the \hi\ disk (at 690, 720 and 900 arcsec) but \citet{hla11} also detected [NII] out to 1140 arcsec or 1.4 D$_{25}$ on the receding side of the galaxy. The most interesting result was that both sets of measurements imply that the RC is declining past the last previously measured \hi\ velocity point \citep{puc91}. From this, \citet{bfq97} conclude that it suggests that the dark halo of NGC 253 may be truncated near the \hi\ edge and provides further support for the link between dark matter and \hi\ \citep{bos78, car89, car90a, car90b, puc90, job90, hoe01, swa12, meu13}. Most importantly, it would provide a means to probe the gravitational potential beyond the edge of the \hi\ disk.

\begin{figure}
\includegraphics[width=\columnwidth]{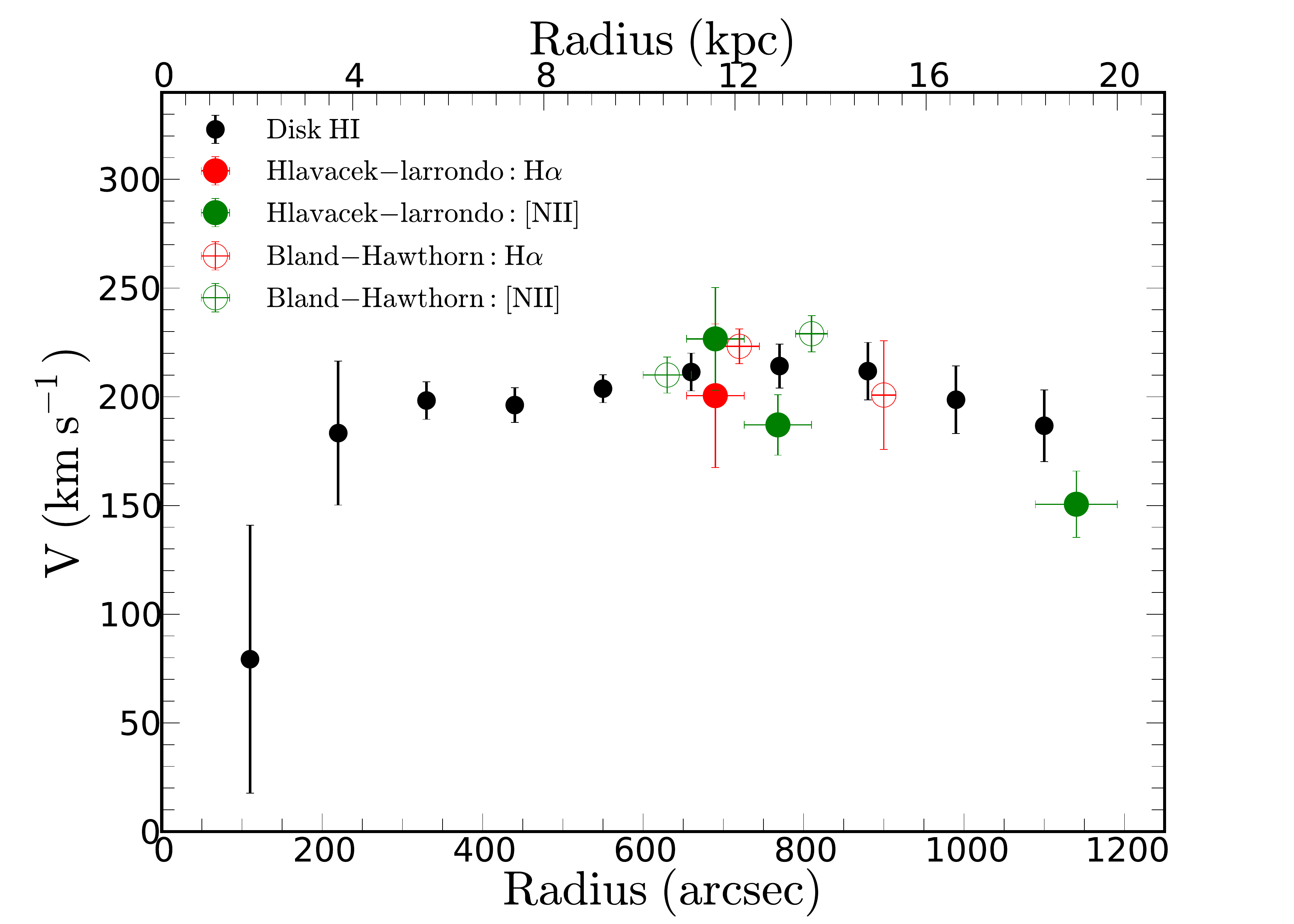}
\caption{Comparison of the \hi\ disk kinematics with the Fabry-Perot data of \protect\citet{bfq97} and \protect\citet{hla11}.}
\label{fig:HI_FP}
\end{figure}

Figure \ref{fig:HI_FP} compares our \hi\ RC to the optical FP velocity measurements on the receding side of NGC 253. Within the errors the \hi\ and optical velocities agree very well and seem to confirm that the RC of NGC 253 declines for R $\ge$12 kpc. Could a bad choice of orientation parameters, mainly $PA$ and/or $i$, mimic a declining RC? As seen in Fig. \ref{fig:disk_rocur} \citep[see also][]{kor95}, there is a suggestion that $PA$ might be increasing for the last two rings. Tilted-ring models, run with those higher values, did not change significantly the derived velocities. However, $i$ could have a more important effect but to bring the last point at the level of the flat part of the RC, an $i \sim 56^o$ would be needed which, as shown in Sec.~\ref{sec:3D}, is excluded.

NGC 253 is the second galaxy in the Sculptor group with a declining RC.  \citet{car90a}  and \citet{deb08} showed NGC7793 to also have a declining RC, which is contrary to what is observed for the vast majority of spiral galaxies.

\subsection{Star formation in NGC 253 using WISE data}
\label{sec:WISE}

If we want to understand the gas flows in the starburst galaxy NGC 253, it is important to understand its star formation.
WISE imaging provides the ideal platform for global property studies of large galaxies, such as NGC 253, due to the large field of view and the broad photometric bands that cover both the stellar (3.4 and 4.6 $\mu$m, or the W1 and W2 bands) and ISM components (12 and 22 $\mu$m, or W3 and W4 bands).   Reconstructions using a "drizzle" sampling technique \citep{jar12} that conserves the native angular resolution (6 arcsec in the 3.4 $\mu$m band) were carried out for NGC 253.   

The resulting mosaics fully cover the galaxy and its immediate environment. However, a significant impediment to the 12 and 22 $\mu$m imaging arises from the nuclear starburst itself, saturating the inner few pixels that comprise the unresolved nucleus.  To recover this lost data, we apply a PSF-technique that was developed by T. Jarrett for saturation in Spitzer IRAC and MIPS imaging (see IRAC Instrument Handbook) and successfully deployed for study of the Circinus Galaxy \citep{for12}.  The rectified WISE imaging of NGC 253 is shown in Figure \ref{fig:WUVXdiskHI} and Figure \ref{fig:WUVXhaloHI}, where all bands are combined into a four-color graphic: W1 assigned shades of blue, W2 green, W3 orange and W4 red.

Deploying the characterization pipeline of the WISE Enhanced Resolution Atlas \citep{jar13}, foreground stars are identified and removed, the two-dimensional shape of NGC 253 is determined and photometry is extracted.  The one-$\sigma$ isophotal (23.24 Vega mag arcsec$^{-2}$ or 26.6 mag arcsec$^{-2}$ in AB) diameter is found to be 42.2 arcmin with an axis ratio of 0.276.  The corresponding integrated flux densities are 12.74, 8.44, 47.95, and 118.04 Jy, for W1, W2, W3 and W4 respectively, indicative of very strong PAH emission (W3) and warm dust continuum (W4) arising from the active star formation. We note that the "total" flux, as inferred from the isophotal plus extrapolated disk emission, is only a few per cent larger, consistent with a relatively truncated stellar disk. 

As demonstrated in \citet{jar13} and \citet{clu14}, the stellar mass-to-light ratio has a simple dependence on the WISE W1 and W2 integrated fluxes.  With our adopted distance of 3.47 Mpc, the corresponding W1 absolute magnitude and Log [in-band luminosity $L_{w1}$ ($L_{\odot}$)] is -24.24 and 10.99 , respectively.  The W1-W2 color is 0.21 mag, and employing the M/L relation from \citet{clu14}, the inferred Log [stellar mass ($M_{\odot}$)] is then 10.33, indicative of the significant (mass) evolved stellar population that represents the "back bone" of NGC 253.  

Star formation activity may be gauged by the WISE 12 and 22 $\mu$m luminosities,  Log [L ($L_{\odot}$)] = 9.67 and 9.77, respectively. Here we apply the relations in \citet{clu14} to estimate the dust-obscured star formation rate (SFR):  4.9 and 5.1 $M_{\odot}$ yr$^{-1}$, respectively based on the 11.3 $\mu$m PAH emission and the 22 $\mu$m dust continuum.  Although this global SFR is typically small compared to luminous infrared galaxies, it is significant in the Local Volume, similar to the starburst M82 and to the barred grand-design spiral M83 \citep{jar13}.   

\begin{figure}
\includegraphics[width=\columnwidth]{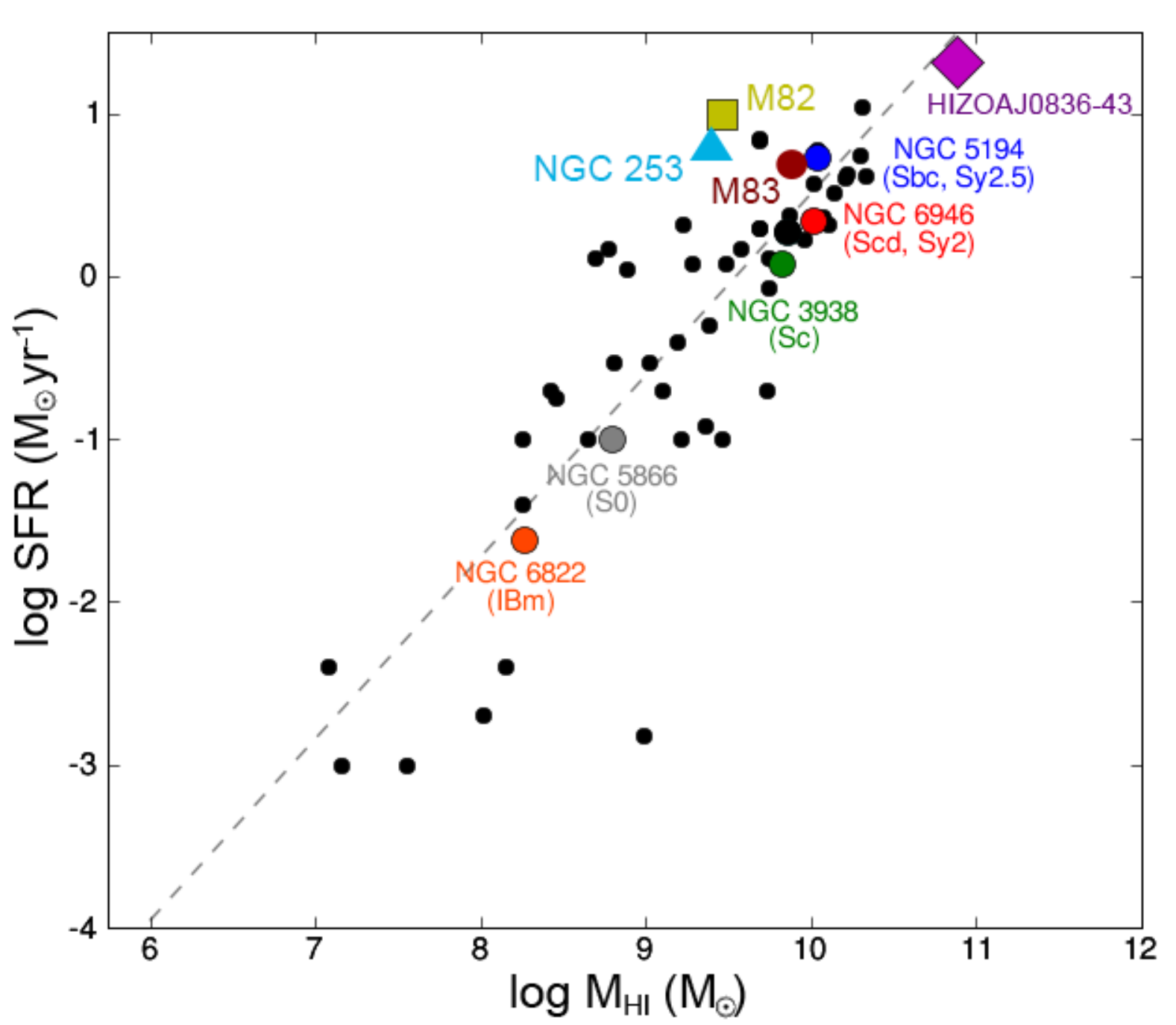}
\caption{Obscured star formation rate compared to the total neutral hydrogen content.  Adapted from \protect\citet{clu10}, the sample is from SINGS \protect\citep{ken03} with highlighted galaxies, notably nearby starbursts M82 and NGC 253 (this work).}
\label{fig:SFRMHI}
\end{figure}

We infer a star formation density of 1.7 M$_{dot}$/year/kpc$^2$ for the central kpc of NGC 253, consistent with a nuclear starburst.
For normal galaxies, \citet{clu10}  showed a clear linear trend in global SFR relative to the \hi\ gas content, effectively the Kennicutt-Schmidt law.  Starburst galaxies, such as M82 and NGC 253, deviate from this relation due to the enhanced nuclear starburst phase, as seen in Figure \ref{fig:SFRMHI}.   Finally, normalizing the SFR with the stellar mass, the resulting Log [specific SFR (yr$^{-1}$)] is -9.86 and  -9.83, respectively W3 and W4.  This disk building rate is relatively high for nearby galaxies -- NGC 253 is still actively building its disk and bulge Ð and is comparable to other large star forming spirals, notably M83 and NGC 6946 \citep{jar13}.

\subsection{Origin of the Anomalous \hi: inflow or outflow ?}
\label{sec:flow}

As discussed in the introduction, at least three possible scenarios could be envisaged to explain the anomalous (halo) gas that is observed. The first one is the classical galactic fountain {\it outflow} 
scenario, where gas is being expelled from the disk through multiple supernova explosions from clusters of massive stars \citep{sha76}. The second one is the {\it inflow} scenario of gas of external
origin in the form of primordial gas clouds left over from the formation of the galaxy  \citep{oor66}. Finally, that halo gas could come from ISM torn out of dwarf galaxies during a close encounter with NGC 253 \citep{put12}.

It is clear from this study that the first scenario can explain a large part of the anomalous gas observed. The known starburst nature of NGC 253, with its SFR $\ge 5$ M$_{\odot}$ yr$^{-1}$ (see Sec.\ref{sec:WISE}) and the large amount of hot
halo gas seen along the minor axis in the X-ray observations \citep{pie00}, all point to gas being expelled from the nuclear region.  This starburst-driven super wind was studied in great detail by \citet{wes11} through deep H$\alpha$ imaging with the MPG/ESO 2.2 m WFI and optical spectroscopy with the VLT/VIMOS-IFU and with the WIYN/SparsePak IFU. They investigate the known minor axis outflow cone, which is well-defined in the H$\alpha$ imaging and kinematics between radii of 280 and 660 pc from the nucleus. Kinematic modelling indicates a wide opening angle ($\sim$60$^0$), an inclination consistent with that of the disc and deprojected outflow speeds of a few 100 \kms\ that increase with distance above the plane.

A dual origin for the gas in the halo in the form of galactic fountains originating within the boiling star-forming disk and a strong galactic super wind emanating from the nuclear starburst, has previously been suggested by \citet{pie00}.
It is thus quite possible that the classical galactic fountain (ionized outflow seen in the H$\alpha$ emission, cooling down through radiative processes and raining back onto the disk) provides most of the extra planar gas up to 5-10 kpc, while the gas higher up in the halo could originates from the starburst region, expelled as hot gas and then cooled to its neutral form.The most convincing piece of evidence that supports the outflow origin for the extra planar gas is seen in
Figure \ref{fig:VELFI}, where it is apparent that the halo gas retained the kinematical signature of the disk. 

As suggested by a referee, the location of the anomalous gas, situated more toward the edge of the disk (see Fig. \ref {fig:WUVXhaloHI}) could possibly be interpreted as the result of the superwind outflow, as in the case of NGC 1482 \citep{Hota05}. It looks as if the anomalous gas in the halo is pushed by this outflow and is providing collimation to the outflowing gas. This halo gas would have a different origin than the outflow gas, most probably through galactic fountains, which is suggested by the observation that it retains traces of the main disk kinematics.

If the anomalous gas was an inflow from external origin, it is expected that those clouds would include material at forbidden velocities.  After a careful visual search of the \hi\ cube of NGC 253, no such clouds with that expected peculiar kinematics were uncovered. However, looking at the size of the inflow clouds, e.g. in NGC 891 \citep{oos07}, of only 1-2\arcmin, such clouds would most probably not be detected in our data cube because of beam dilution (beam $\simeq 3.5\arcmin \times 3\arcmin$). Thus, we cannot exclude completely the presence of such small inflow clouds. Finally, our data seems also to exclude gas torn off a dwarf galaxy during a close encounter since again we would not necessarily expect that gas to have the same kinematics as that of the disk. 

The last question left to answer: where spatially is most of the detected anomalous gas? Looking at Fig. \ref{fig:C_disk_ano}, it can be seen that the halo gas has a much shallower gradient than the disk gas and is lagging by $\sim$ 100 \kms\ in the outer parts. Modelling done \citep[see e.g.][]{fra05} of extra planar gas predicts a lag of $\sim$15 \kms\ kpc$^{-1}$, consistent with observations of edge-on galaxies. Applied to NGC 253, this suggests that the bulk of the gas should be around 6-8 kpc from the plane of the main \hi\ disk. If it were closer to the disk  \citep[see e.g.][]{hes09}, it would have a steeper gradient in the inner parts and a much smaller lagging velocity.

\subsection{Origin of the starburst nature of NGC 253}
\label{sec:starburst}

One of the principal reasons often given for high SFR is galaxy-galaxy interaction \citep[see e.g.][]{lt78}, whether in the form of merger, close encounter or interaction with the IGM. A good example is the M81 group where the three core members M81, M82 and NGC 3077 are known to be closely interacting from \hi\ studies \citep{vdh79, yun94} and, at the same time, all three show evidence for an AGN or starburst activity, most likely induced by the on-going tidal interactions.

In contrast, the environment of NGC 253 appears much more devoid of any sign of gravitational interaction either with low-mass companions or the IGM. The search of \citet{cot91} for dwarf galaxies in the Sculptor group (see their Fig. 4) do not show any obvious nearby dIrr with \hi\ emission, while a few possible dSphs candidates were identified. As for the presence of a significant IGM component, the study by \citet{put03} suggests that the surrounding region has very little intergalactic gas (see their Fig. 12). 

However, the apparent lack of evidence of any actual interaction does not tell us anything about possible past interactions of NGC 253 with its environment. For example, as shown by \citet{whi99}, the spin of the NGC 253 disk is consistent with torquing from NGC 247, which is its closest ($\sim 350$ kpc) large companion.
Moreover, both galaxies appear to have truncated \hi\ disks \citep{puc91, car90b}, since  they barely extend out to their optical diameter, This is contrary to what is generally seen in late-type spirals where the \hi\ disk is usually much more extended than their stellar disk \citep{san83}. As an example, the \hi\ disk of NGC 300, another late-type spiral in the Sculptor group, has a diameter more than 1.5 times the optical diameter \citep{puc90}.

Despite its apparent isolation, there is some evidence which suggests that NGC 253 may have been recently involved in a merger. For example, in a recent X-ray study to spatially resolve the starburst region of NGC 253 \citep{wik14}, the 4-6 keV data clearly shows what appears to be a double nucleus (see their Fig. 4). Earlier studies \citep{ana96, pra98} also found evidence of two dynamically distinct systems close to the centre of NGC 253. As discussed by \citet{dav10}, one system has an axis of rotation that differs from that of the galaxy disk, while the other appears to be counter-rotating with respect to the galaxy disk. These are all signatures of a recent merger. \citet{das01} estimate that the merger may have involved the accretion of 10$^6$ M$_{\odot}$ of material $\sim$10$^7$ years ago. In view of the actual isolation of NGC 253, this is the most likely explanation for the starburst activity in its nucleus.

\section{Summary and Conclusions}
\label{sec:con}

This study presents the results of $\sim$150 hours of KAT-7 observations in \hi\ line mode, of which $\sim$115 hours were spent on NGC 253. For the analysis, the data were smoothed to a velocity resolution of 5 \kms\ and the spatial resolution (synthesized beam) is 213\arcsec  x 188\arcsec.  As much as 33\% more flux was detected by KAT-7, compared to previous VLA observations for a total \hi\ mass of $2.1 \pm 0.1$ $\times 10^{9}$ M$_{\odot}$. A sensitivity limit of $\sim$1.0 $\times$ 10$^{19}$ cm$^{-2}$ in column density allows us to uncover a large quantity of extra planar gas out to projected distances from the plane of $\sim$9-10 kpc in the centre and 13-14 kpc at the edge of the disk. However, in the radial direction, the \hi\ disk is relatively small for a late-type spiral, similar to the optical and mid-IR disk, consistent with the truncation scenario. 

A robust method was developed to separate the regular and anomalous \hi\ components in NGC 253. This method, which used
interactive profile fitting in combination with PV diagrams, allowed to fit separately the different components pixel by pixel. For the disk, the RC derived from the KAT-7 data agrees well with the VLA data out to $\sim$12 kpc (the last measured VLA velocity point). For  R $>$ 12 kpc, the RC declines by $\sim$ 20 \kms\ out to the last point at $\sim$18 kpc. This decline, observed in the outer parts, agrees with previous optical FP observations. As for the halo gas, which extends out to $\sim$24 kpc, its RC has a very shallow gradient in the inner parts and is kinematically lagging the disk gas by $\sim$100 \kms in the outer parts. That extra planar cold gas component is seen at the edge of the hot gas component revealed by the X-ray observations. 

Our results point towards an {\it outflow} scenario for the origin of the extra planar gas.
This outflow has two origins. First, as clearly seen in X-ray observations, a lot of hot gas is expelled in the halo from the nuclear starburst region. 
 As the gas expands, it eventually converts back to \hi\ as it cools through radiative losses; raining back down on the disk and feeding subsequent star formation. 
 Secondly, another part of that extra planar gas has most likely a galactic fountain origin from the star-forming disk.
 
The KAT-7 observations presented in this work clearly show that despite its relatively small size
(7 x 12 m antennae), this telescope occupies a niche for detecting large scale low surface brightness diffuse emission over
the $\sim$1$^{\rm o}$ FWHM of its antennae. It should be kept in mind that this telescope was built primarily as 
a testbed for MeerKAT and the SKA such that any scientific result that can be obtained is a bonus.
While most of the extragalactic \hi\ sources would be unresolved
by the $\sim$4\arcmin\ synthesized beam, many projects such as this one and the project on NGC 3109 \citep{car13} 
can be done on nearby very extended objects such as Local Group galaxies or galaxies in nearby groups like Sculptor. 

\section*{Acknowledgments}

We thank the entire SKA SA team for allowing us to obtain scientific data during the commissioning phase of KAT-7. 
The work of CC an TJ is based upon research supported by the South African Research Chairs Initiative (SARChI) of the Department of Science 
and Technology (DST),  the Square Kilometre Array South Africa (SKA SA) and the National Research Foundation (NRF).
The research of DL, EE \& TR have been supported by SARChI, SKA SA fellowships.

\label{lastpage}

\end{document}